\documentclass{article}       % onecolumn (second format)
\usepackage{amssymb}
\usepackage{amsmath}
\usepackage[normalem]{ulem}
\usepackage{comment}
\usepackage{xcolor}
\usepackage{authblk}

\usepackage{xurl}
\usepackage{comment}

\usepackage{geometry}
\usepackage{graphicx}
\usepackage{pdflscape}
\usepackage{hyperref}

\begin{document}
% Prepared for Advances in Statistical Analysis
% https://www.springer.com/journal/10182/updates/17805238

\title{\textcolor{black}{Waiting for Dabo:\footnote{With apologies to Samuel Beckett.} A machine learning model for predicting Power 4 college football coaching hire success
} 
%\sout{What does not get observed can be used to make age curves stronger: estimating player age curves using regression and imputation} \thanks{This material is based upon work supported by the U.S.~National Science Foundation under Grant No. CNS-1919554}
}
%\subtitle{I don't think\\we have a subtitle}

%\titlerunning{Waiting for Dabo}        % if too long for running head
\author[1]{Michael Schuckers}
\author[2]{Austin Hayes}
\affil[1]{Professor of Data Science and Sports Analytics\\ 

School of Data Science, UNC Charlotte\\

\href{mailto:schuckers@charlotte.edu}{schuckers@charlotte.edu}}
\affil[2]{Data Science Undergraduate\\ UNC Charlotte}


\date{17 November 2025}
\maketitle

\begin{abstract}

\textcolor{black}{ Using data  on 103 recent P4 college football hires, we built a statistical model for predicting a coach’s success at their new school.   For each hire, we collected data about their background and experiences, the previous success as a head coach or coordinator and their success since hiring.  Over 50 variables on these factors were recorded though we used 29 of these in building our predictive model.  Our measure of success is based upon Bill Connelly’s SP+ team ratings relative to the performance on the same metric of the school in the 15 year prior to their selection as head coach.  Using a cross-validated regularized linear regression, we obtain a predictive model for coaching success.  Among the important factors for predicting a successful hire are having been a previous college head coach, leaving a job as an Offensive Coordinator, age and quality of the hiring school's team in the previous 15 years. While we do find these factors are important for the prediction of a successful coaching hire, the trends here are weak.  With 66\% accuracy, the model does identify coaching hires that will outperform team performance in the 15 years before the hire.  However, no combination of these factors leads to high predictability of identifying a successful coaching hire.  All of the data and code for this paper are available in a Github repository.}

\end{abstract}

%\sout{The impact of age on athlete performance has received attention across sport.
%First, we highlight how selection bias is linked to the ages in which ages we observe players perform. This approach is used to generate underlying distributions of how players move in and out of sport organizations. Second, motivated by methods for missing data, we propose novel estimation methods of age curves by using both observed and unobserved (imputed) data.}

\section{Introduction}
College football is big business.  The 
changes due to the use of player's name, image and license (NIL) that allow
players to be openly (and appropriately) paid for their services, has brought
a new influx of funds.  A December 2024 study by CNBC using fiscal year 2023 numbers
estimated that the top university athletic programs are valued at over $1$ billion US
dollars and have annual revenue of over $200$ million, \cite{cnbc_college_worth}.  
Unsurprisingly the executive that leads the college football operation,
the head coach, are well compensated.
In a 2023 study, David Evans found that
``80\% of the Highest-Earning Public Employees are College Head Coaches.'' \cite{highest_state} 

Being
a successful college football head coach requires multifaceted talent in 
player development, in gameday decision making, in player evaluation and in 
recruitment of players.  Consequently, the selection of a new coach for top 
universities is an important one.  The analysis develops a machine learning
predictive model which finds that there are
a handful of factors that predict a coaching hire's success.  However,
it is clear that our model's predictions are far from certain and that
there is a great deal of noise relative to the signals in the data
about who would make a successful coach.

In the rest of this paper we look at past hiring
of college football head coaches to build a preditive model of the 
characteristics and factors that influence their success.  
The next Section introduces the data that we collected through
both automated and
manual means.  Summaries of the variables we used as predictors are given in Section 3.
We describe our model building process and results in Section 4.  Finally, we
conclude with a discussion of our results.

\begin{figure*}[htbp]
\centering
\caption{Plots for selected categorical predictors vs SP Rel}
\label{fig:cat1}

\begin{minipage}[htbp]{0.45\textwidth}
    \includegraphics[width=\textwidth]{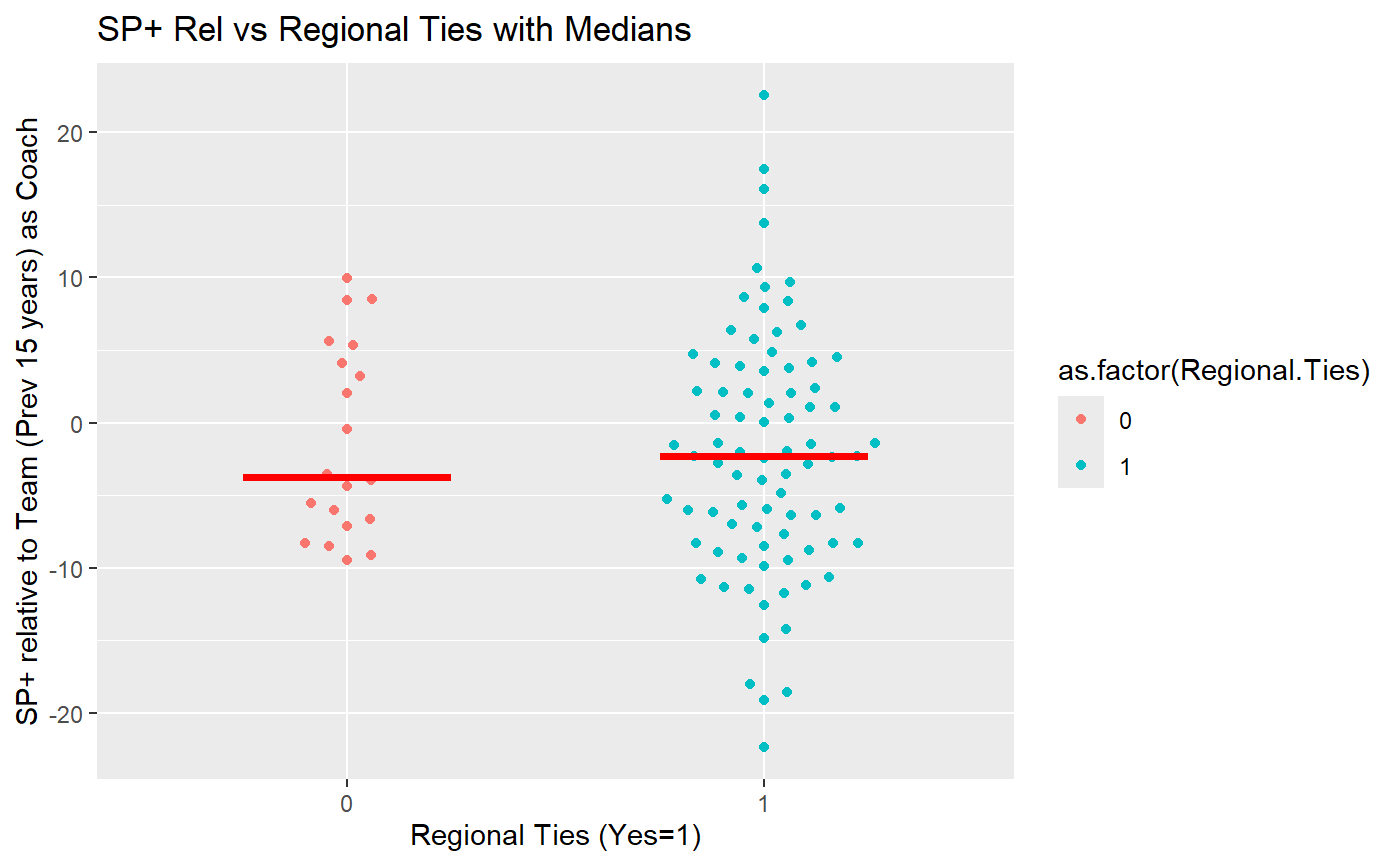}
    %\caption{
       \hspace*{0.1in}
       {\footnotesize (a) Regional Ties}
    \end{minipage}
\begin{minipage}[htbp]{0.45\textwidth}
    \includegraphics[width=\textwidth]{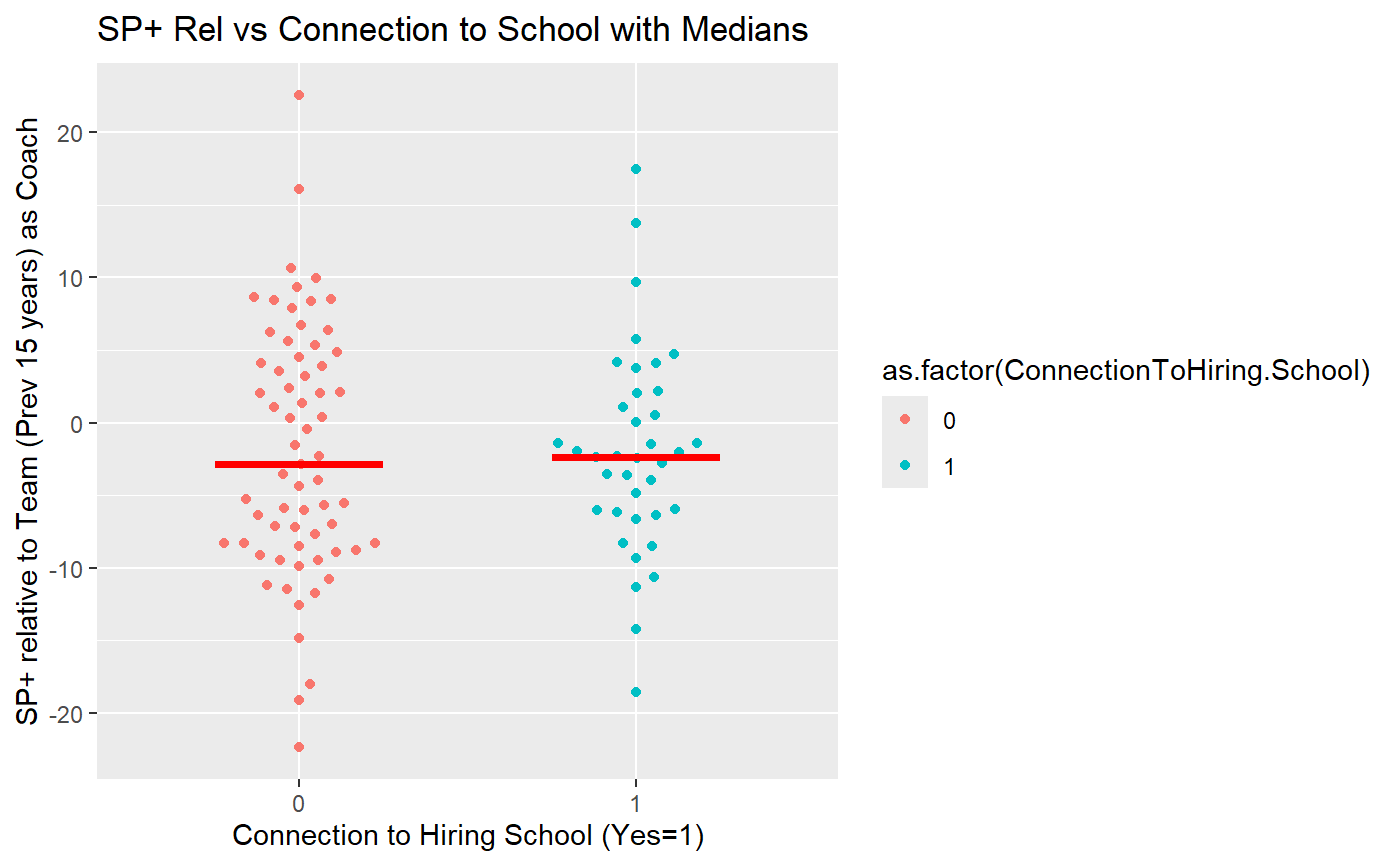}
    %\caption{
       \hspace*{0.1in}
       {\footnotesize (b) Prior Connection to School}
    \end{minipage}
\hfill 

\vspace{0.1in}
\begin{minipage}[htbp]{0.45\textwidth}
    \includegraphics[width=\textwidth]{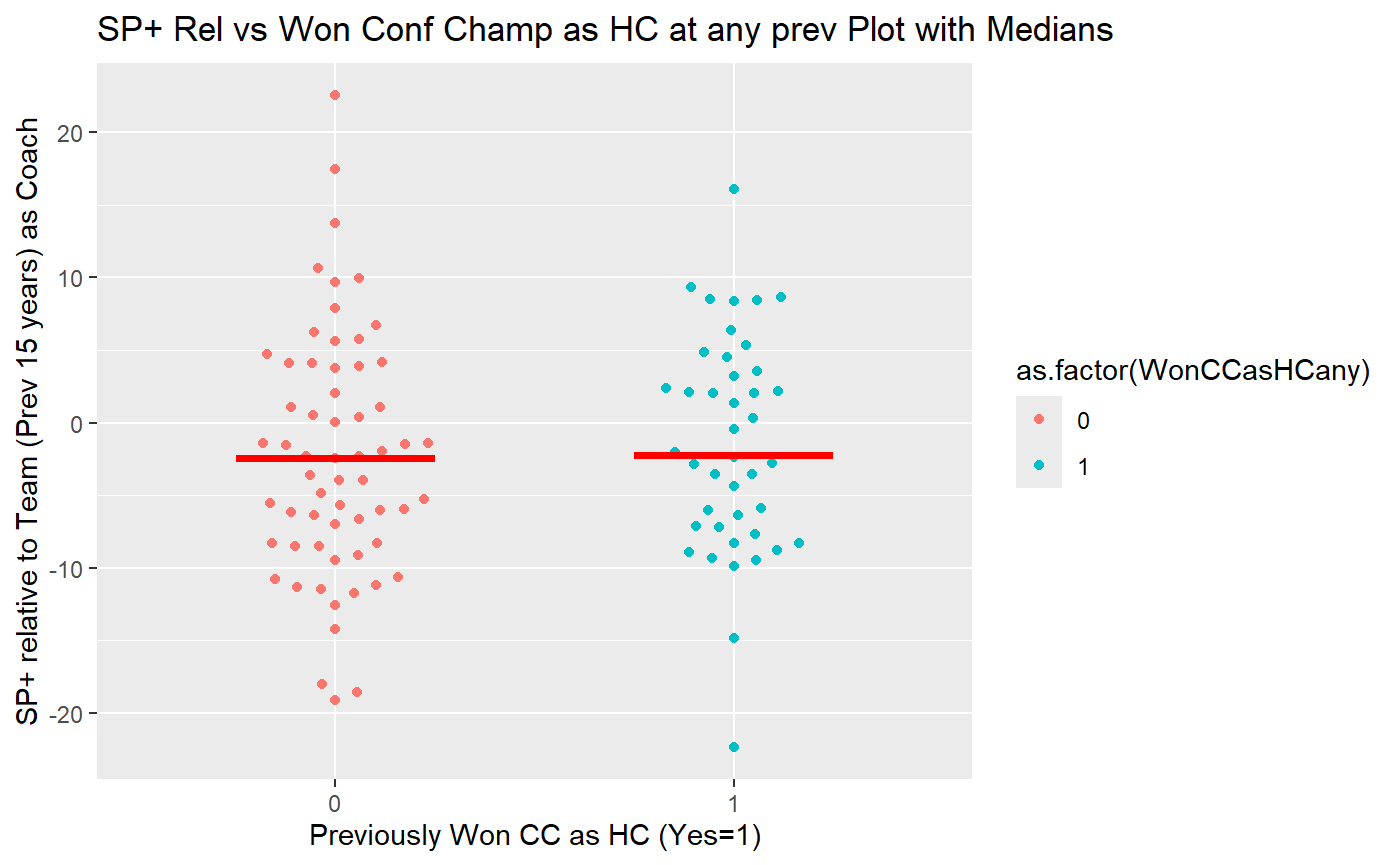}
    %\caption{
       \hspace*{0.1in}
       {\footnotesize (c) Previously won Conf. Champ. as HC}
    \end{minipage}
  %\hfill
 \begin{minipage}[htbp]{0.45\textwidth}
    \includegraphics[width=\textwidth]{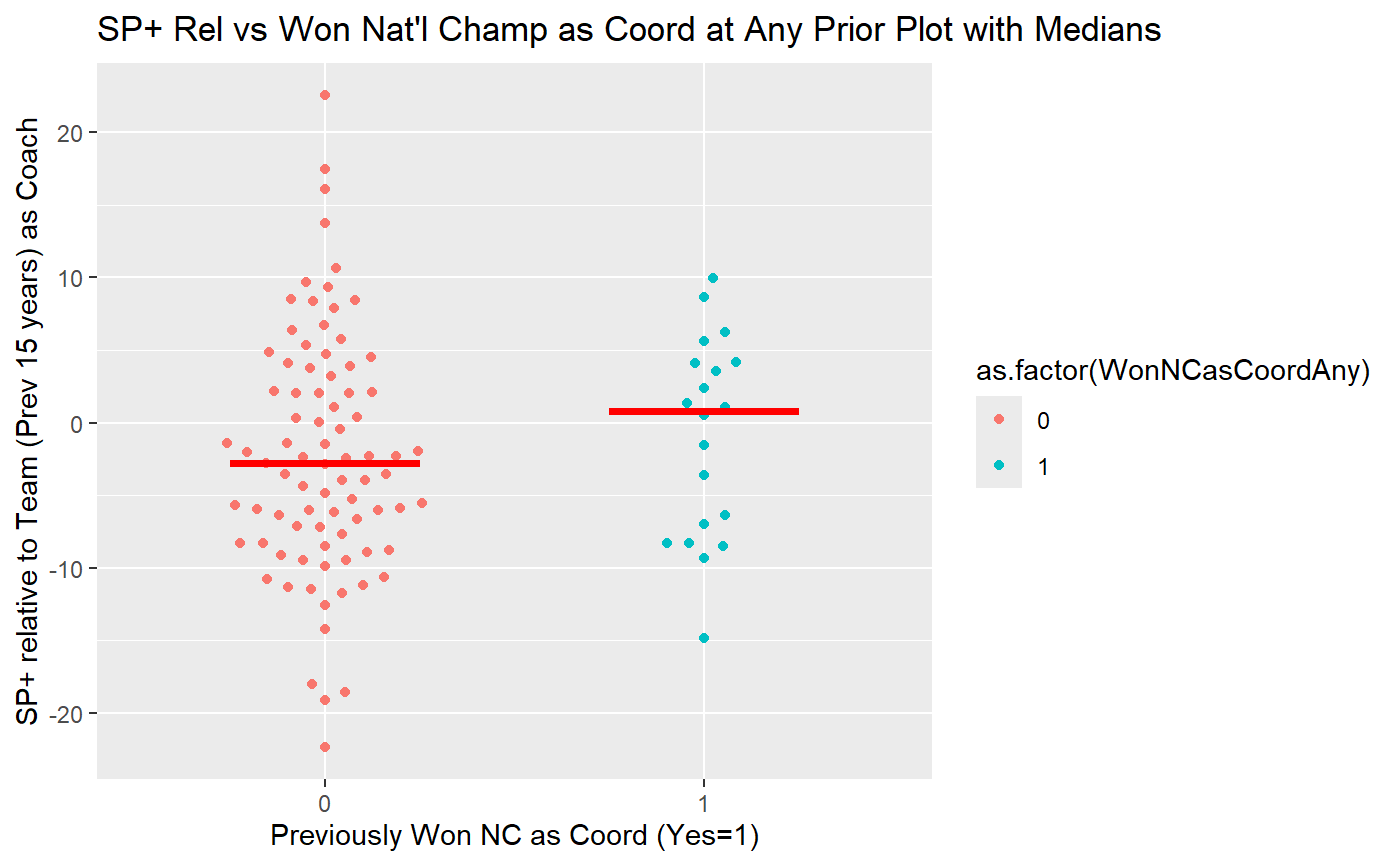}
    %\caption{
       \hspace*{0.1in}
       {\footnotesize (d) Previously won Natl. Champ. as Coordinator}
    \end{minipage}
\hfill

\vspace{0.1in}
\begin{minipage}[htbp]{0.45\textwidth}
    \includegraphics[width=\textwidth]{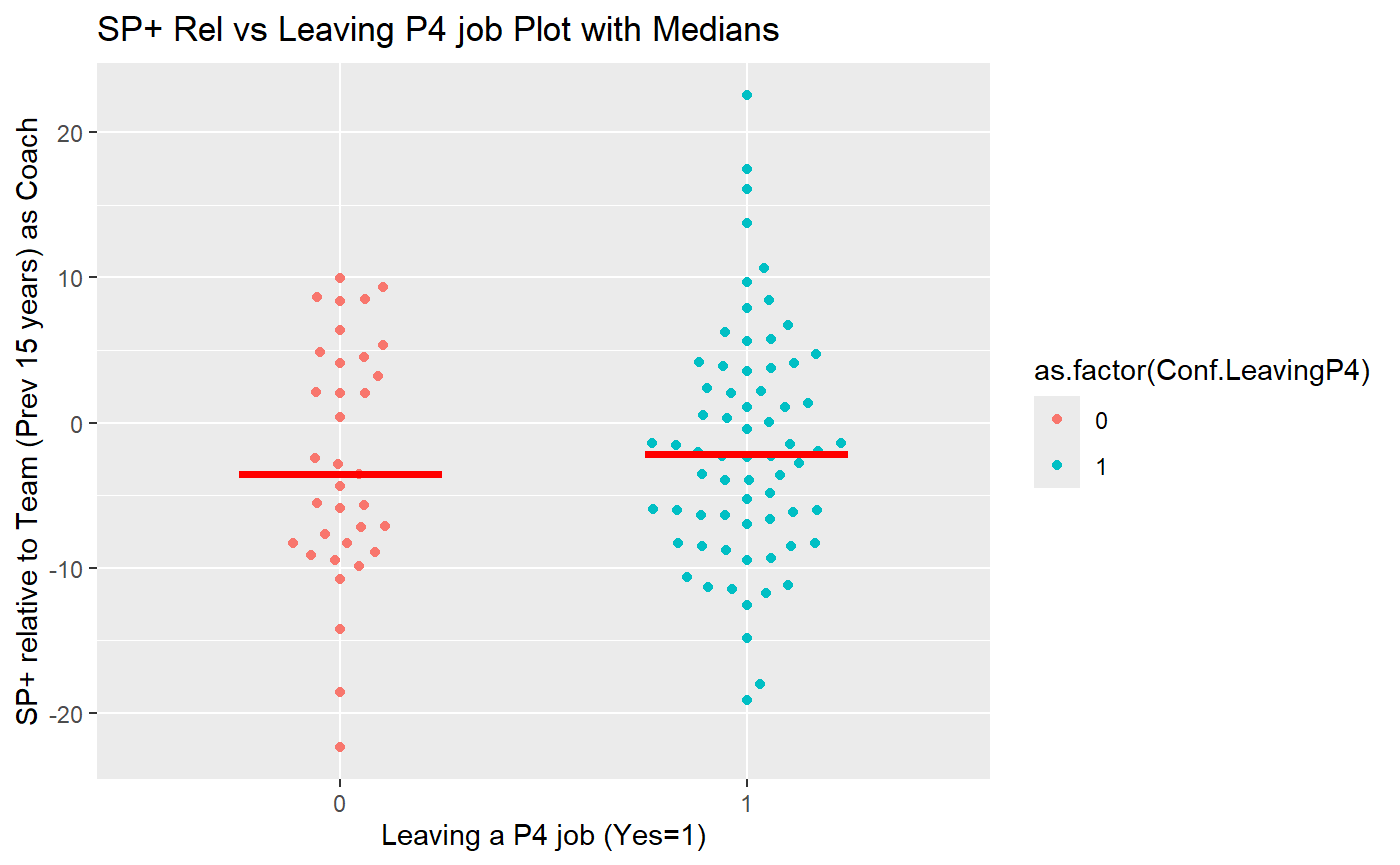}
    %\caption{
       \hspace*{0.1in}
       {\footnotesize (e) Previous Job with P4 School}
    \end{minipage}
  %\hfill
 \begin{minipage}[htbp]{0.45\textwidth}
    \includegraphics[width=\textwidth]{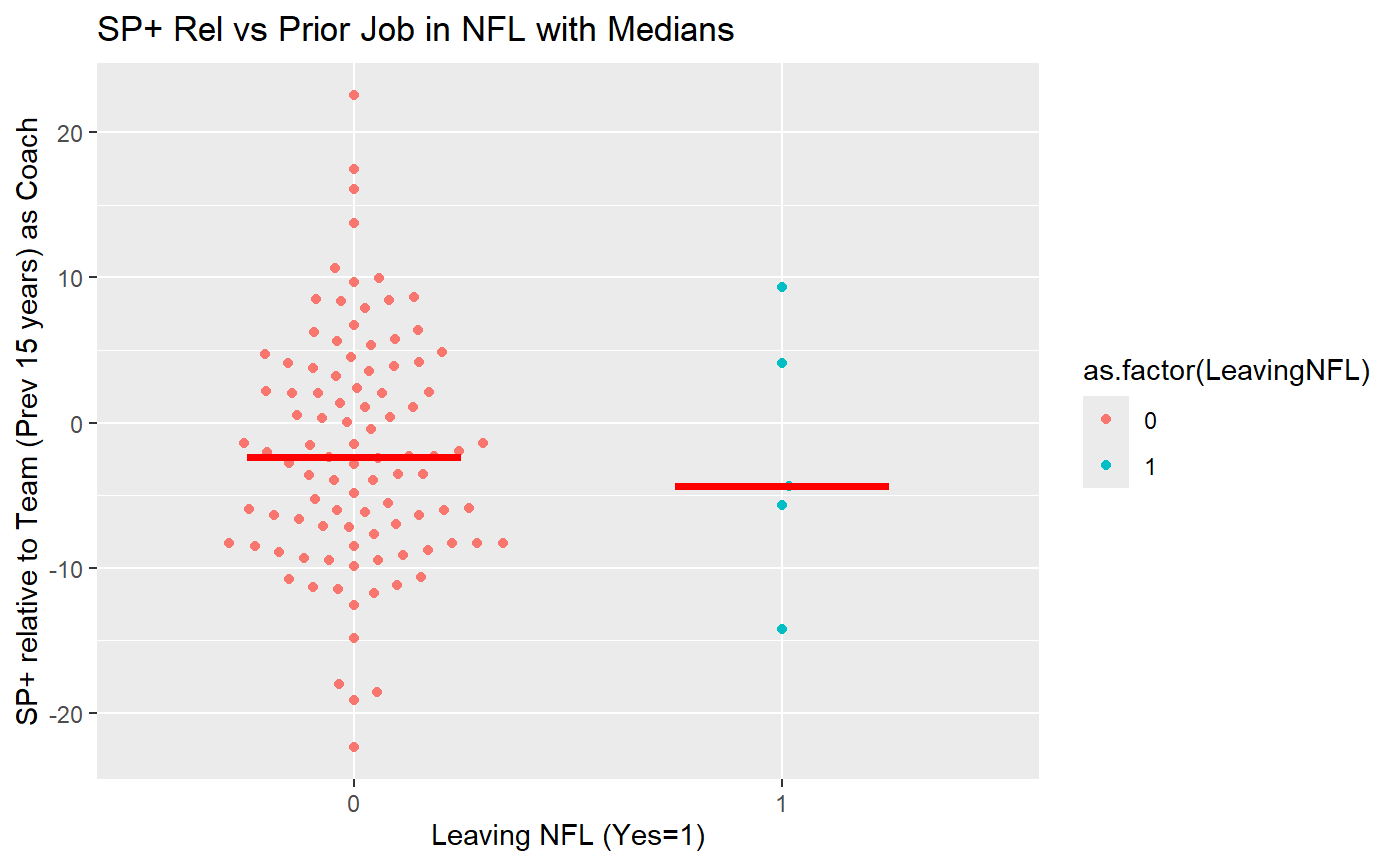}
    %\caption{
       \hspace*{0.1in}   
       {\footnotesize (f) Previous Job with NFL}
    \end{minipage}
\hfill

\vspace{0.1in}
\begin{minipage}[htbp]{0.45\textwidth}
    \includegraphics[width=\textwidth]{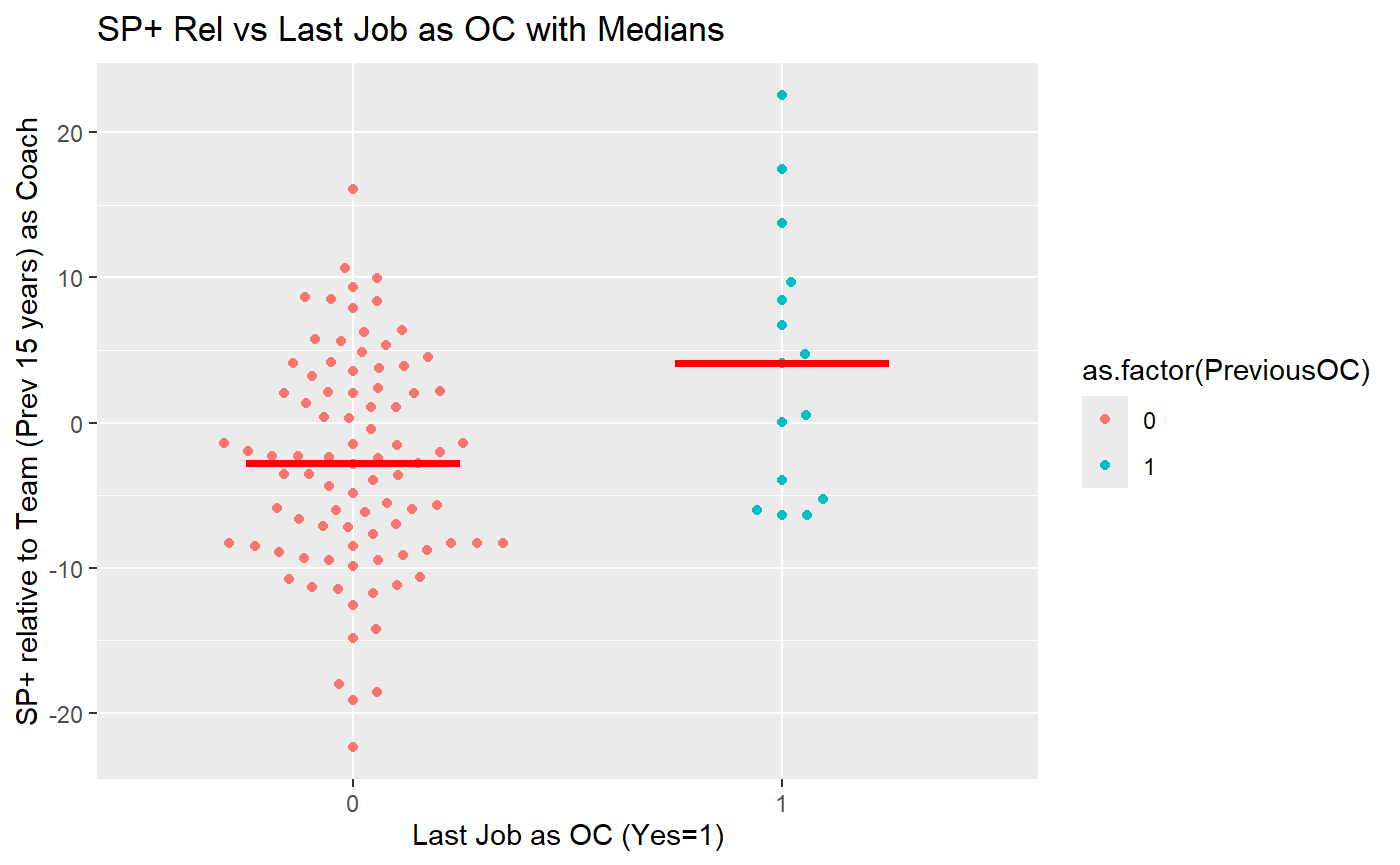}
    %\caption{
       \hspace*{0.1in}
       {\footnotesize (g) Previous Job as OC}
    \end{minipage}
  %\hfill
 \begin{minipage}[htbp]{0.45\textwidth}
    \includegraphics[width=\textwidth]{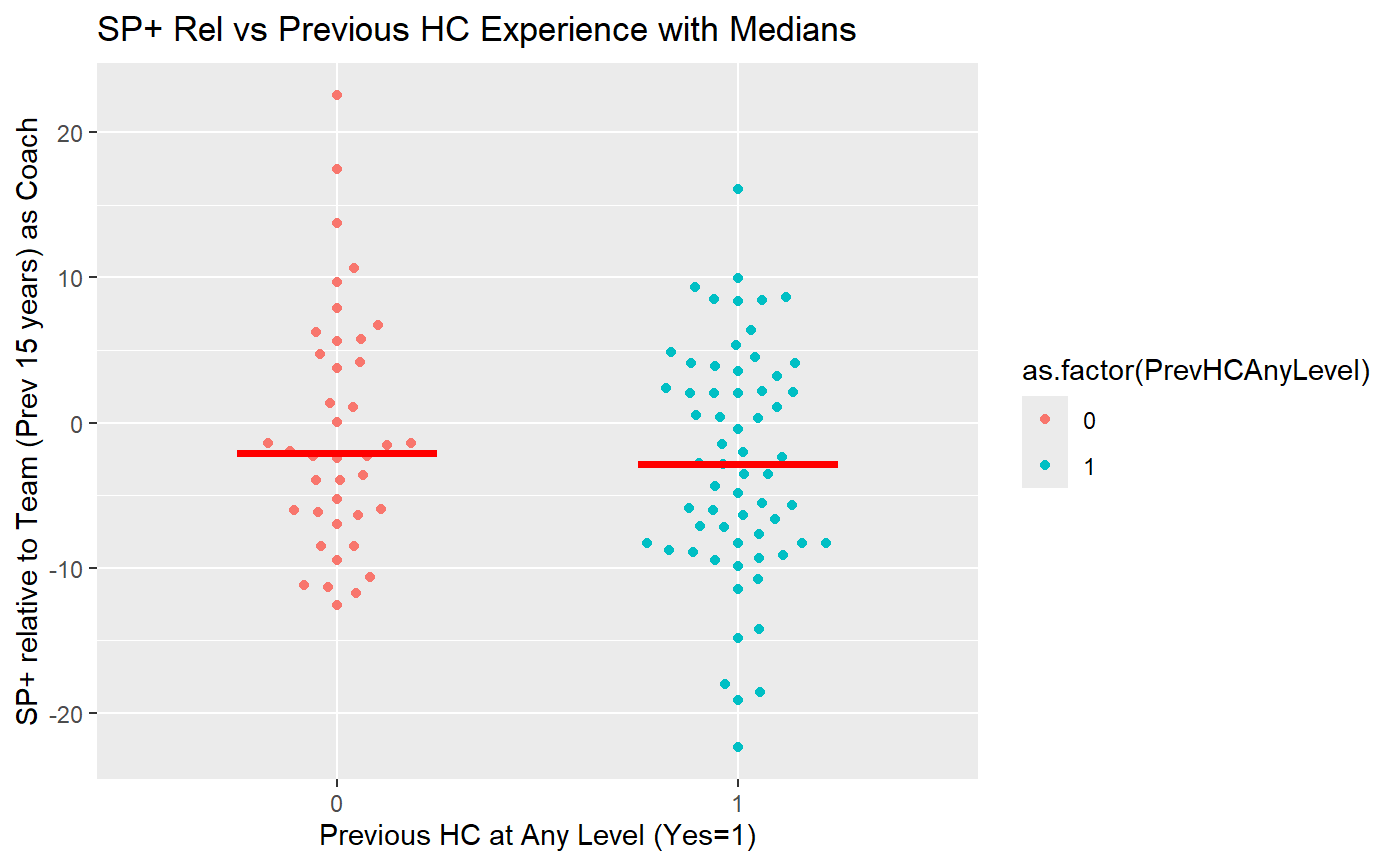}
    %\caption{
       \hspace*{0.1in} 
       {\footnotesize (h) Previously been HC any level college}
    \end{minipage}
\hfill

\end{figure*}

\section{Data}
The data set used for this analysis is a collection of Power 4 plus Notre Dame (P4) coaching hires.  The P4 conferences are the Atlantic Coast Conference (ACC), Big 12, Big Ten and Southeastern Conference (SEC). We began with all coaching hires who started coaching in 2016 or since.  We added at least once coach from each school so that Kirk Ferentz is in our data.  That gave us 120 hires.
To ensure we had sufficient data on how a 
Specifically, we include P4 coaching hires who started coaching between 2016 and 2023.  As our observational units, we chose hires rather than coaches since it is the outcome of the hiring process we are trying to maximize.  Consequently, Kirk Ferentz, the current dean of college football, is in the data.  Nick Saban is not among our data, but James Franklin is in the data, as the only Penn State coach over this time period.  Matt Rhule appears two times because of being hired at Baylor (in 2017) and at Nebraska (in 2023).  We did not include interim coaches and, of course, as the title implies Dabo Swinney is one of our observations. In total there are 103 coaching hires in these data.  

For each hire in our data, we recorded over 50 measurements though we used 29 as predictors below.  Generally these fall into three categories: background variables, previous coaching performance, and performance metrics in the job for which they were hired. Included in the background variables are information about their coaching career, their previous coaching positions, whether or not they have previously coached in the NFL, whether or not they had previously been a HC, whether they had played or coached previously at the hiring institution, etc.  Previous coaching performance variables are recorded and include previous performance as a college head coach.  Variables on a coaches 
background included features such as whether or not they had a previous connection to the hiring school, whether they played P4 football, and whether they had coached in the NFL. A full list of all variables is presented in Table \ref{table:var} of the Appendix.  One of the metrics we considered was having a connection to the hiring school by being an alumni or having played there or having coached there.  Having an alumni as your head coach was previously analyzed by Josh Planos of FiveThirtyEight and found to not impact coaching success either way, \cite{alumni_538}.    To account for regression to the mean, a team that was underperforming relative to historical averages before a coach arrived, we also included the team's average performance for the five years before they arrived. 
\begin{figure*}[htbp]
\centering
\caption{Plots for selected numeric predictors vs SP Rel}
\label{fig:numeric1}

\begin{minipage}[htbp]{0.45\textwidth}
    \includegraphics[width=\textwidth]{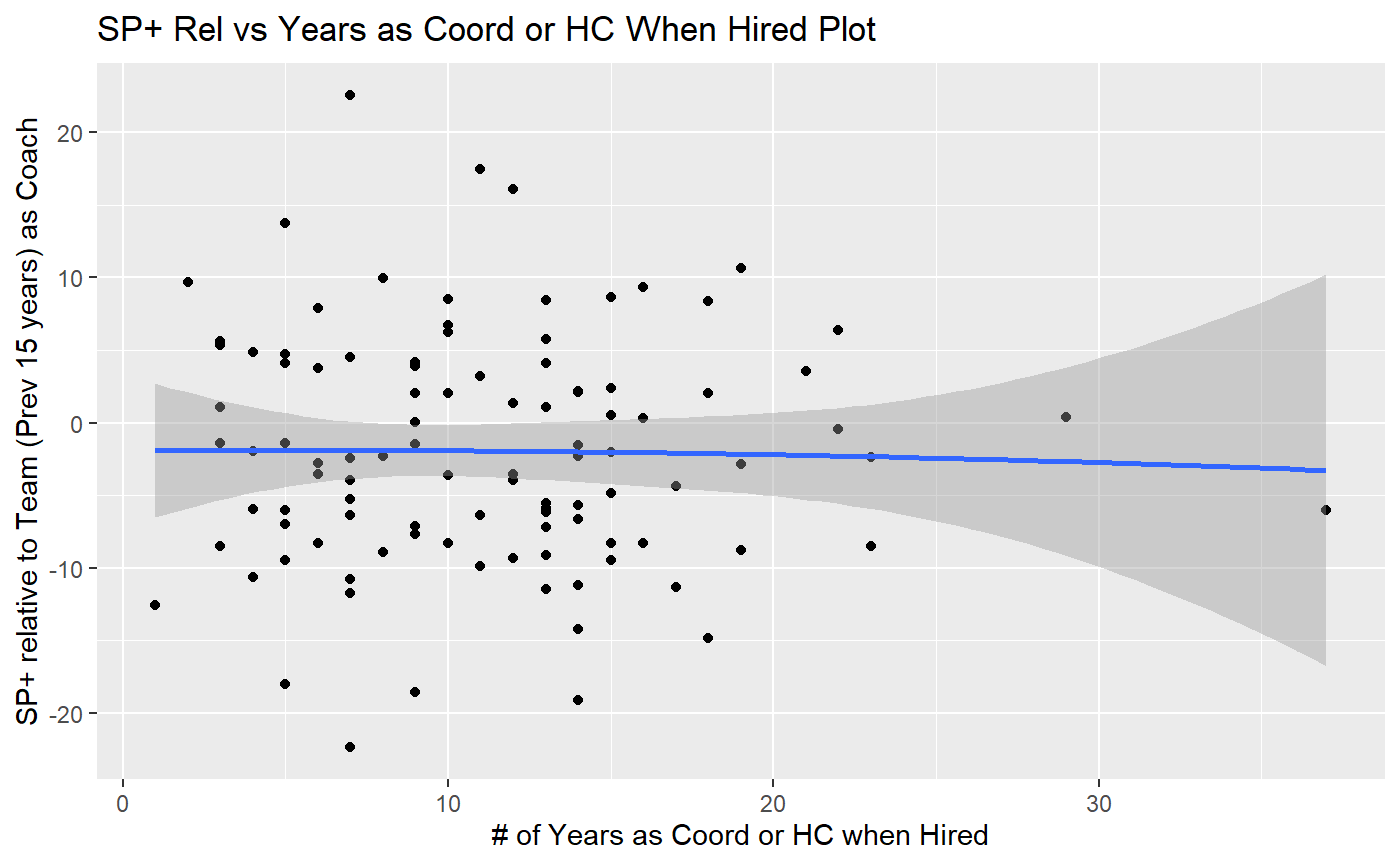}
    %\caption{
       \hspace*{0.1in}
       {\footnotesize (a) Numb. years as Coord or HC}
    \end{minipage}
\begin{minipage}[htbp]{0.45\textwidth}
    \includegraphics[width=\textwidth]{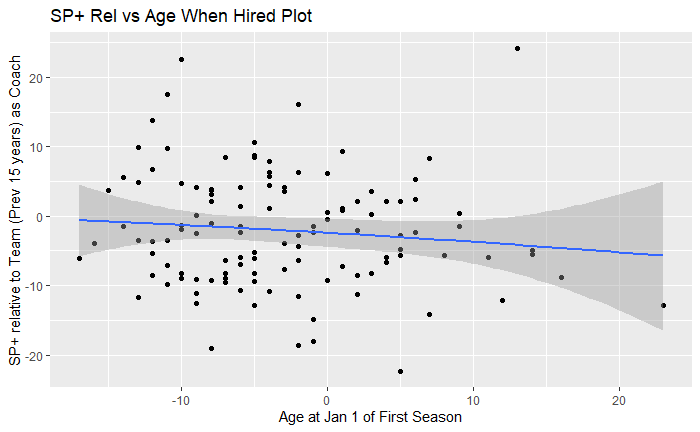}
    %\caption{
       \hspace*{0.1in}
       {\footnotesize (b) Age - 50}
    \end{minipage}
\hfill 

\vspace{0.1in}
\begin{minipage}[htbp]{0.45\textwidth}
    \includegraphics[width=\textwidth]{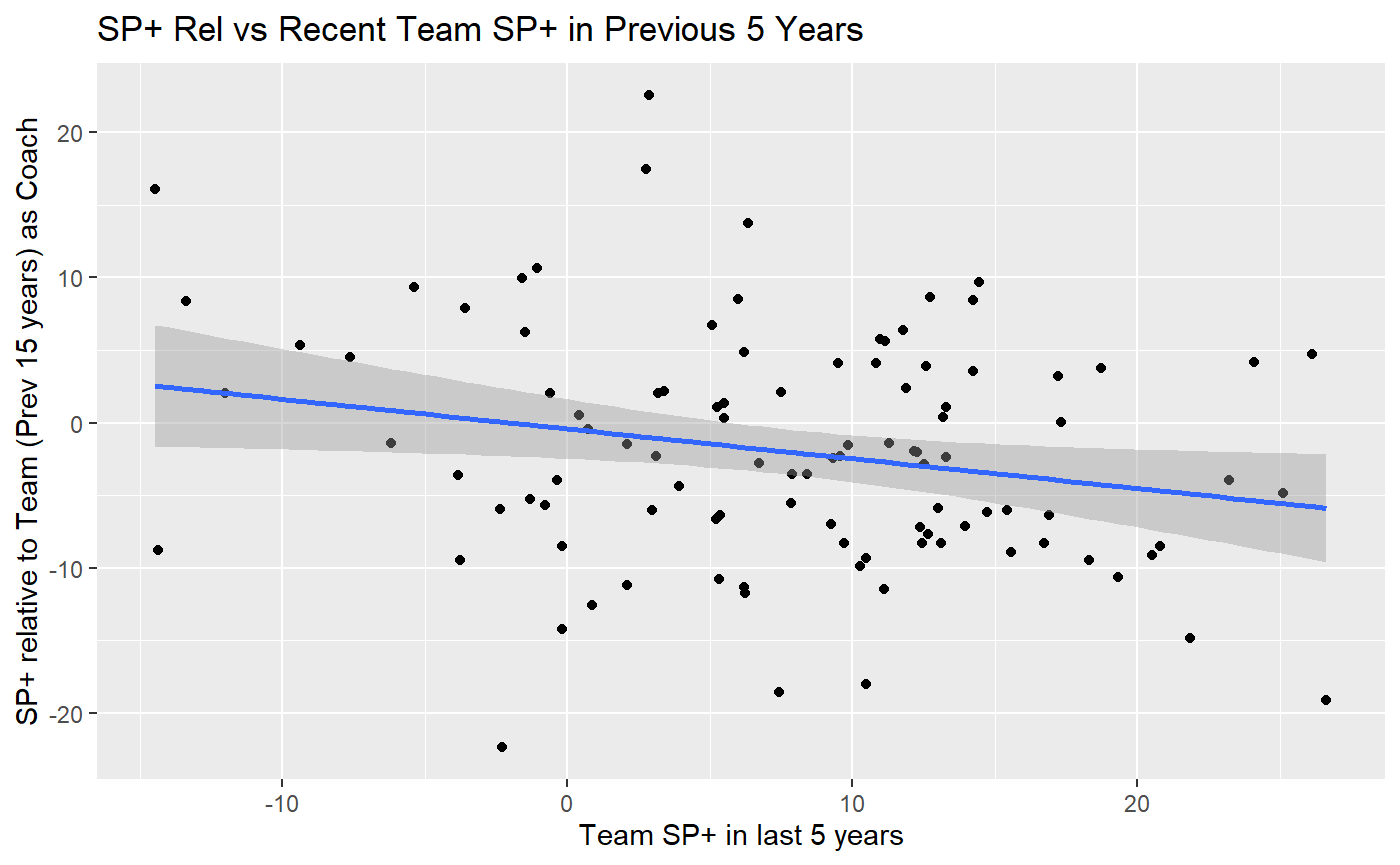}
    %\caption{
       \hspace*{0.1in}
       {\footnotesize (c) Teams SP+ in last 5 years}
    
    \end{minipage}
  %\hfill
 \begin{minipage}[htbp]{0.45\textwidth}
    \includegraphics[width=\textwidth]{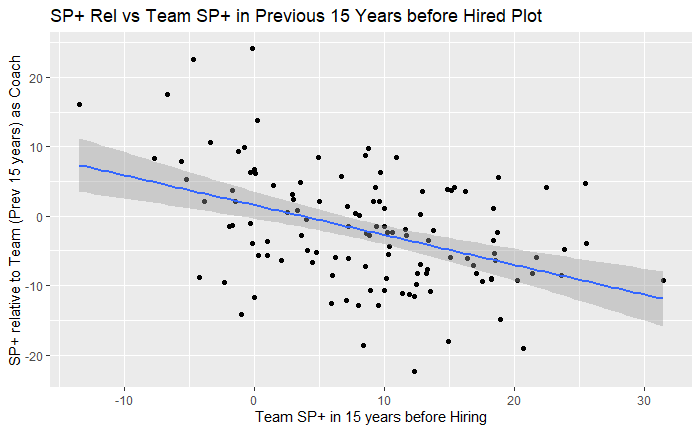}
    %\caption{
       \hspace*{0.1in} 
       {\footnotesize (d) Teams SP+ in last 15 years}
    
    \end{minipage}
\hfill

\end{figure*}

To determine how successful each coach was as a hire we calculated their performance as a HC relative to the historical performance of the university where they were hired.  For performance we considered three measurements.  First we considered a hire's win percentage since winning is arguably the most important outcome for a head coach.  Second, we considered a coach's average 
Simple Rating System (SRS).  This metric, SRS, is a “weighted scoring margin” where the weighted part is based upon the quality of the teams that have been played \cite{intro_srs}.  Third, we considered a coaches average relative SP+ which is a predictive metric developed by Bill Connelly \cite{intro_sp} that is a “tempo- and opponent-adjusted measure” of team efficiency \cite{sp_2025}.  The historical values for win percentage, SRS and SP+ were obtained using the \texttt{cfbfastr} package \cite{cfbfastr}. We augmented those data with wins and losses from the 2025 season.
\begin{table*}[h]
\centering
\caption{Best Performing Coaching Hires by \textit{Sp+ Rel}.}
\label{table:topcoaches}
\begin{tabular}{|l||c|r|}
\hline
Name& Hiring School& \textit{SP+ Rel}\\
\hline
Josh Heupel	&UCF	&22.6		\\
Rhett Lashlee&	SMU	&17.5		\\
Sonny Dykes	&SMU	&16.1		\\
Mike Gundy&	Oklahoma State&	13.8	\\	
Mike Elko&	Duke&	10.7		\\
Matt Campbell	&Iowa State&	9.9\\		
Dabo Swinney	&Clemson&	9.7		\\
Bret Bielema	&Illinois&	9.5		\\
Lane Kiffin	&Ole Miss&	8.8		\\
Josh Heupel	&Tennessee&	8.7		\\
\hline
\end{tabular}
\end{table*}

One of the significant challenges for this type of analysis is to define success.  This is certainly the case here, and  in no small part is due to the expectations for each school.  What might be a successful record at Arizona, might not be a success at Alabama.  What gets a coach an extension in College Park, might be a reason to be fired in College Station or State College.  Consequently, our definition of coaching performance is relative (to team history) rather than absolute.
For relative performance, the mean \footnote{\texttt{mean(, trim=0.1) in R}} of the middle $80\%$ of years of the performance metric for their time as a head coach and subtracted the team average of that same metric for the fifteen years before the coach was hired. We chose
to use SP+ as our measure and we refer to \textit{SP+ Rel} as our outcome variable
below.   Note that the correlation between SRS relative to past team performance and SP+ relative to past team performance was 0.95, results using the former as a response are likely to be very similar to those for the latter and so we did not consider models with different responses.  The correlation between \textit{SP+ Rel}
and a winning percentage relative to past teams was $0.82$.

Using the R package \texttt{cfbfastr} , we obtained historical team data including
wins and losses through the 2024 season as well as SRS and SP+ ratings into the 2025 season.  
We manually added wins and losses for the 2025 season.  All of these data can be found at:\\ 
\href{https://github.com/schuckers/cfb_coach_analytics}{https://github.com/schuckers/cfb\underline{~}coach\underline{~}analytics}.  Data 
on coaching performance is complete through midday on November 13, 2025.

\section{Data Summaries}

Among these 103 college football head coaching hires in our data, there were 50 who left a job as head coach (HC) at another school, and 63 of our hires had been a HC at some level previous to their being hired.
Twenty-eight of the hired coaches had been Defensive Coordinators (DCs) and 15 were Offensive Coordinators(OCs).  The other previous positions included four position coaches, an offensive analyst, four Associate or Assistant HCs and one hire who was not employed in football.  While more coaches were hired directly from DC positions, 40 of the 103  hires (39\%) had defense as their background.  Sixty-three of the hires had been a head coach at the college or NFL previously and in total only 24 had previously coached in the NFL.  Thirty-one percent of the coaches were hired to jobs within the same conference.  Only 16
of our hires were promoted from the coaching staff of the same institution including Dabo Swinney.  The average age of our coaches when they were hired is approximately 46 years.  

The majority of hired coaches, 65, had not worked for, played at or attended the school that hired them.  While the vast majority of the hired coaches played college football, just under 50\% of them, 49, played for a P4 school and only ten played in the NFL.  

Below when we refer to recent metrics we will be talking about a five year period and when we refer to previous metrics we will be talking about a 15 year period.

\begin{figure*}[h]

\centering
\caption{Additional plots of categorical predictors vs \textit{SP+ Rel}
\label{fig:append:cat}}

\begin{minipage}[htbp]{0.4\textwidth}
    \includegraphics[width=\textwidth]{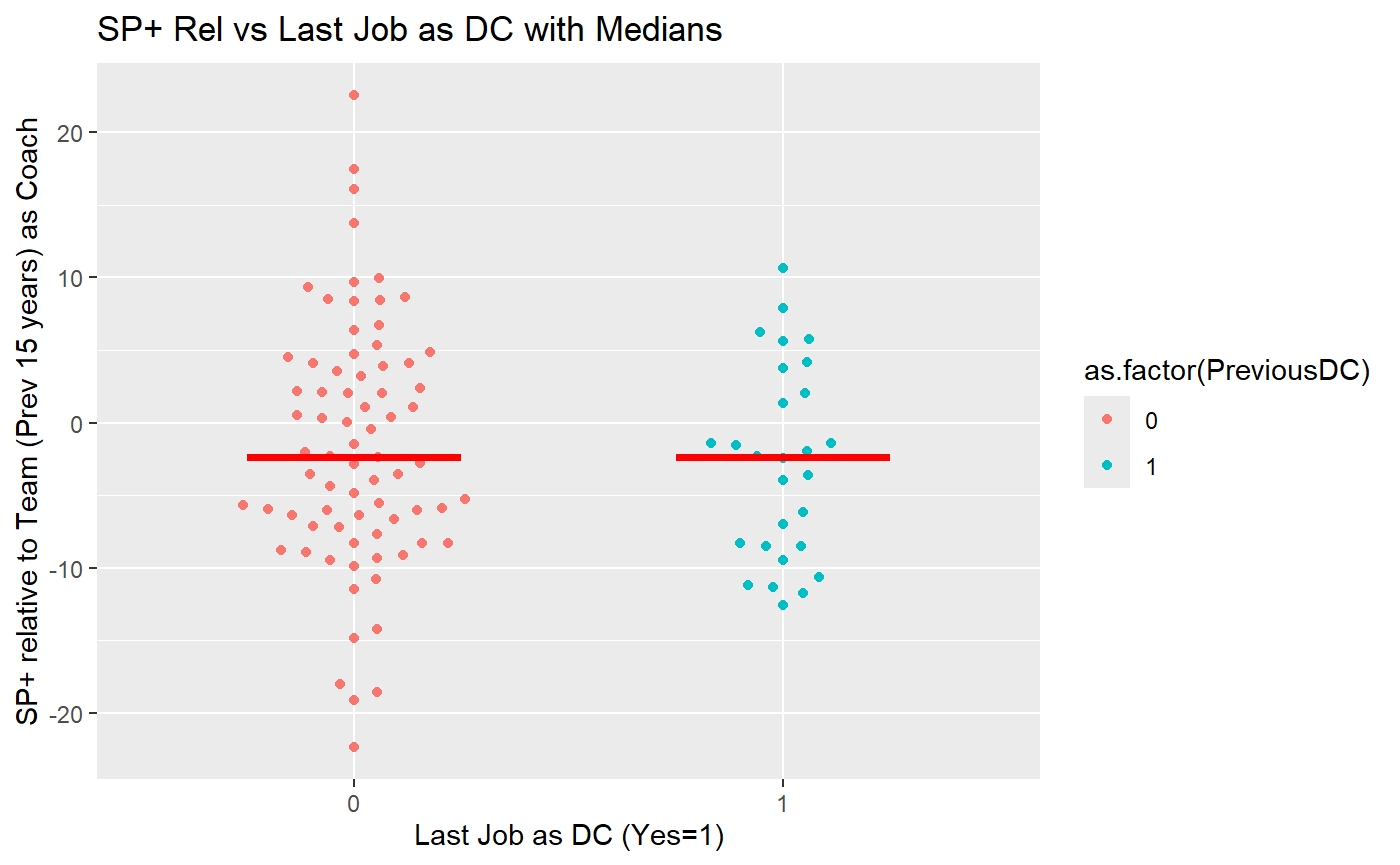}
    %\caption{
       \hspace*{.1in}
       {\footnotesize (a) Previous job as DC}
    \end{minipage}
  %\hfill
\begin{minipage}[htbp]{0.4\textwidth}
    \includegraphics[width=\textwidth]{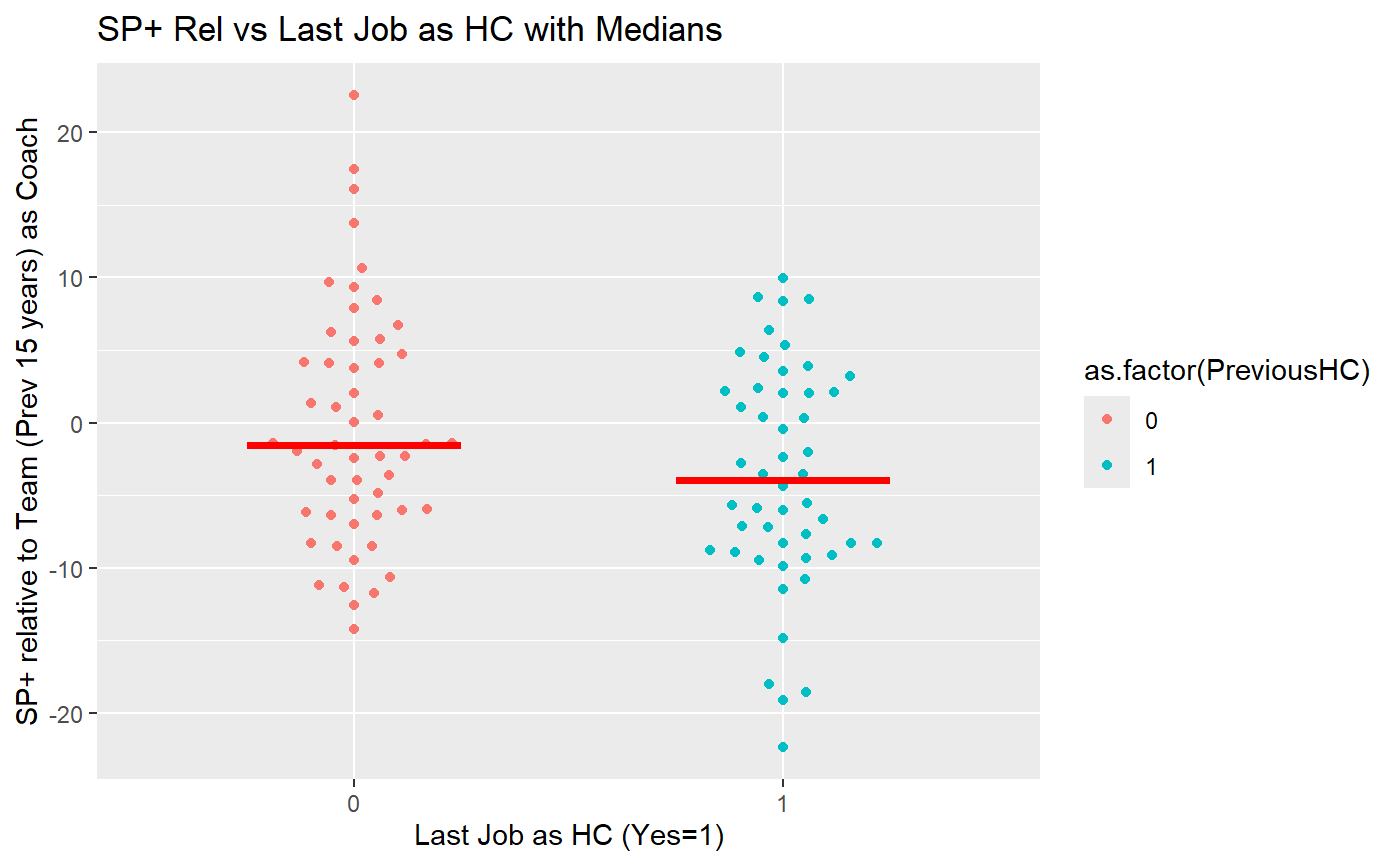}
    %\caption{
       \hspace*{.1in}
       {\footnotesize (b) Previous job as HC}
    \end{minipage}
\hfill 

\vspace{0.2in}

\begin{minipage}[htbp]{0.4\textwidth}
    \includegraphics[width=\textwidth]{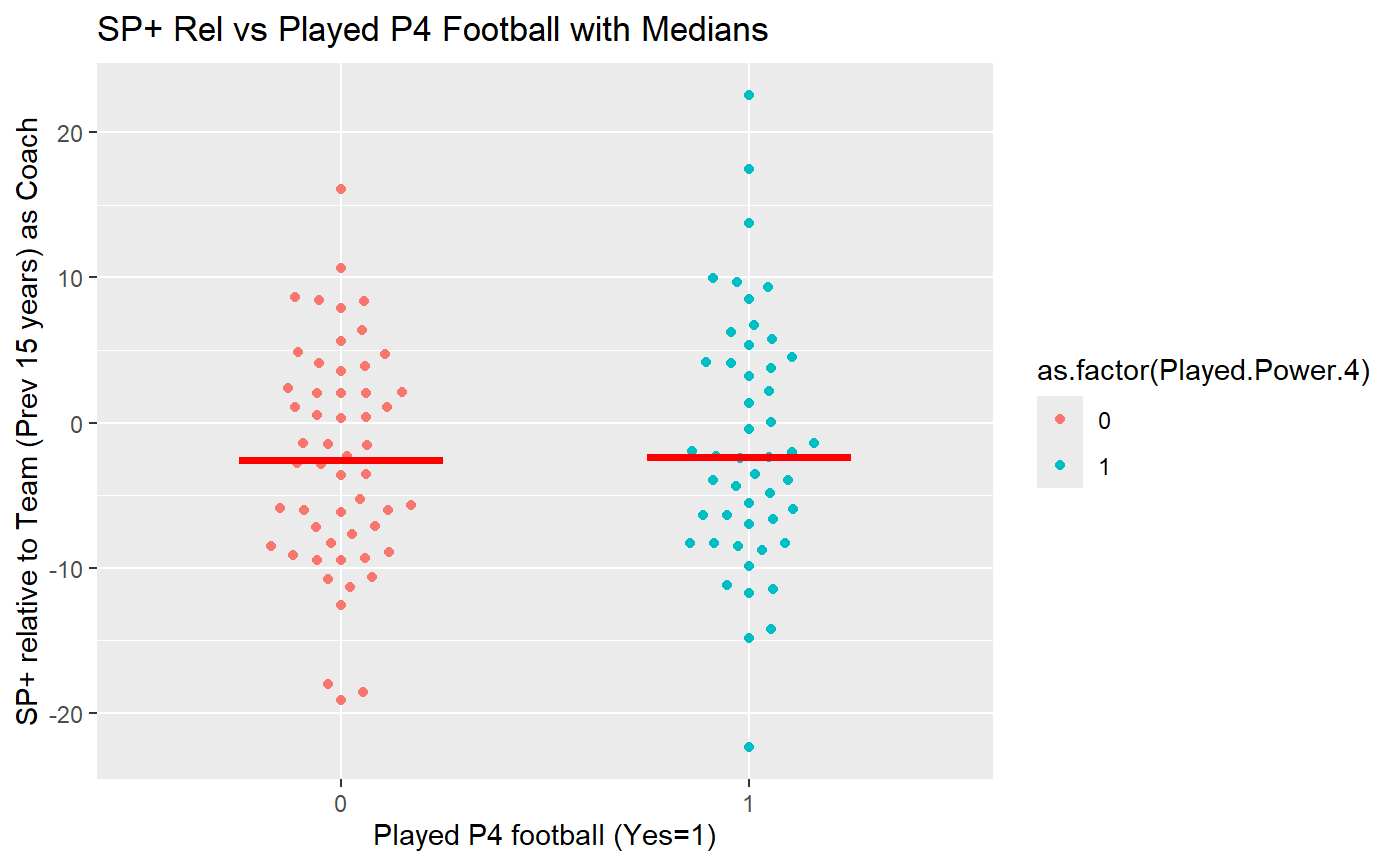}
    %\caption{
       \hspace*{.1in}
       {\footnotesize (c) Played P4 football}
    \end{minipage}
  %\hfill
 \begin{minipage}[htbp]{0.4\textwidth}
    \includegraphics[width=\textwidth]{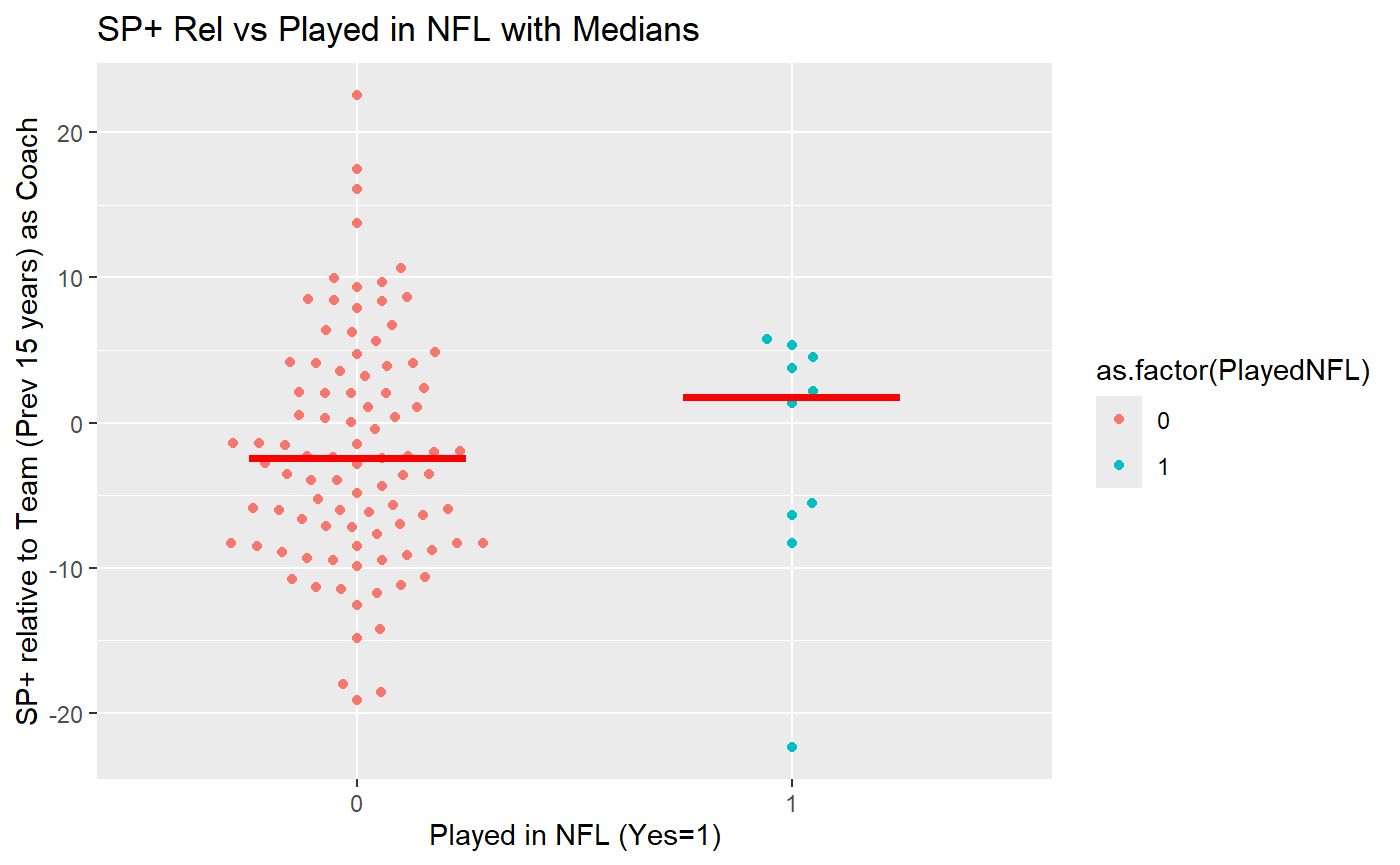}
    %\caption{
       \hspace*{.1in}
       {\footnotesize (d) Played in the NFL}
    \end{minipage}
\hfill

\begin{minipage}[htbp]{0.4\textwidth}
    \includegraphics[width=\textwidth]{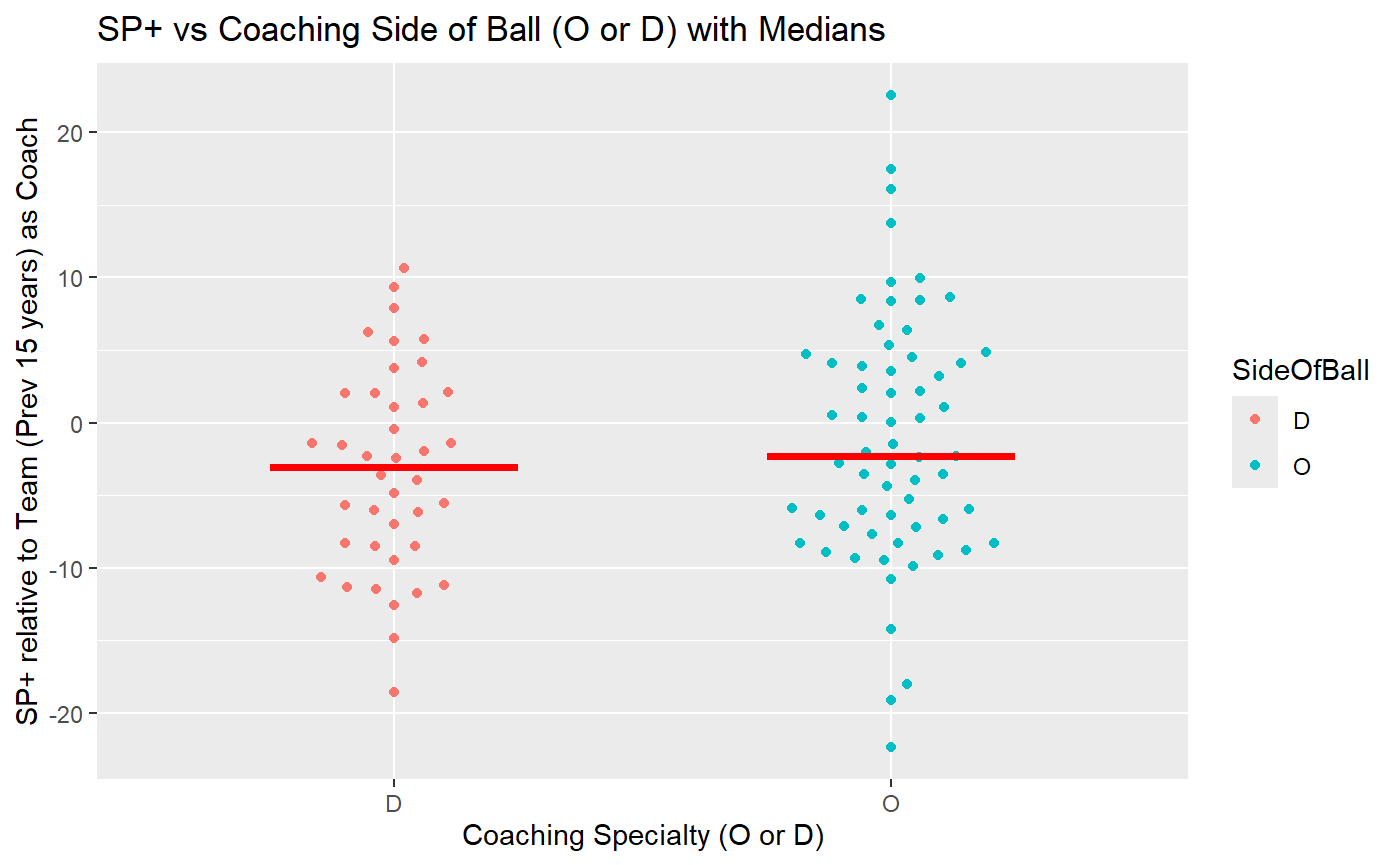}
    %\caption{
       \hspace*{.1in}
       {\footnotesize (e) Side of the Ball}

    \end{minipage}
    \begin{minipage}[htbp]{0.4\textwidth}
    \includegraphics[width=\textwidth]{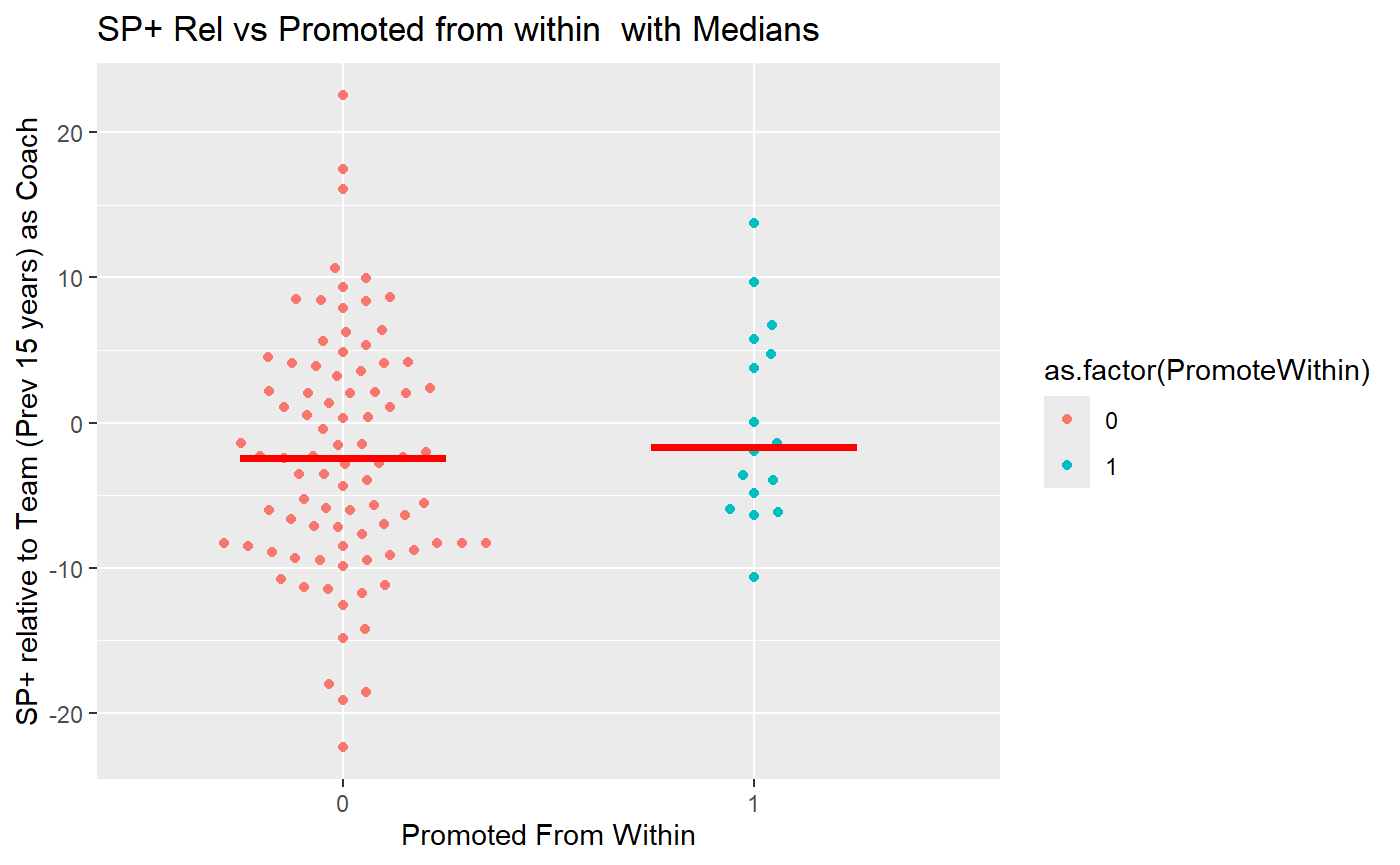}
    %\caption{
       \hspace*{.1in}
       {\footnotesize (f) Promoted from Within}
    \end{minipage}
\hfill
    
\end{figure*}

\begin{figure*}[!t]

\centering
\caption{Additional plots of categorical predictors vs \textit{SP+ Rel}
\label{fig:append:cat2}}

\begin{minipage}[htbp]{0.4\textwidth}
    \includegraphics[width=\textwidth]{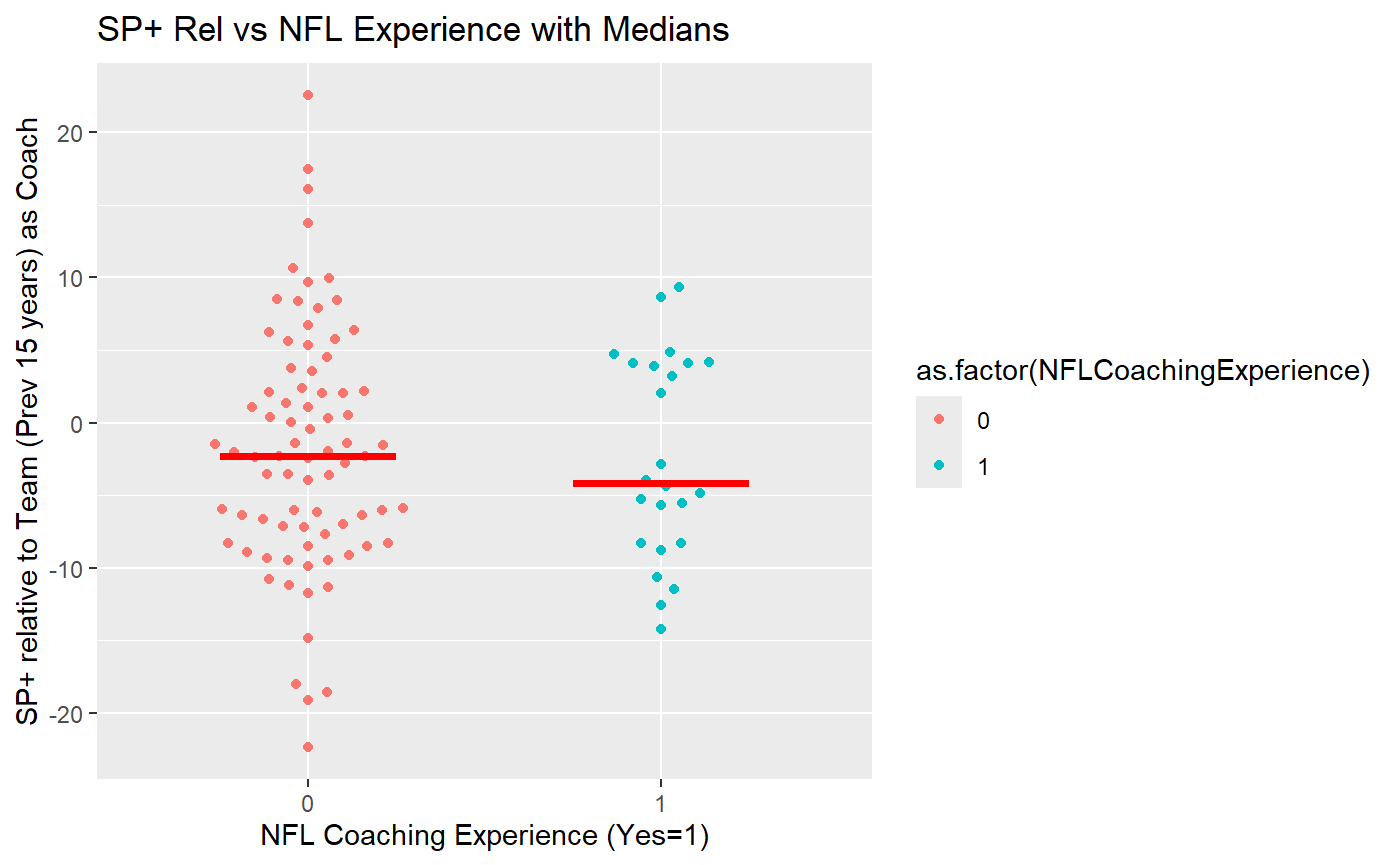}
    %\caption{
       \hspace*{.1in}
       {\footnotesize (a) Has NFL coaching experience}
    \end{minipage}
  %\hfill
\begin{minipage}[htbp]{0.4\textwidth}
    \includegraphics[width=\textwidth]{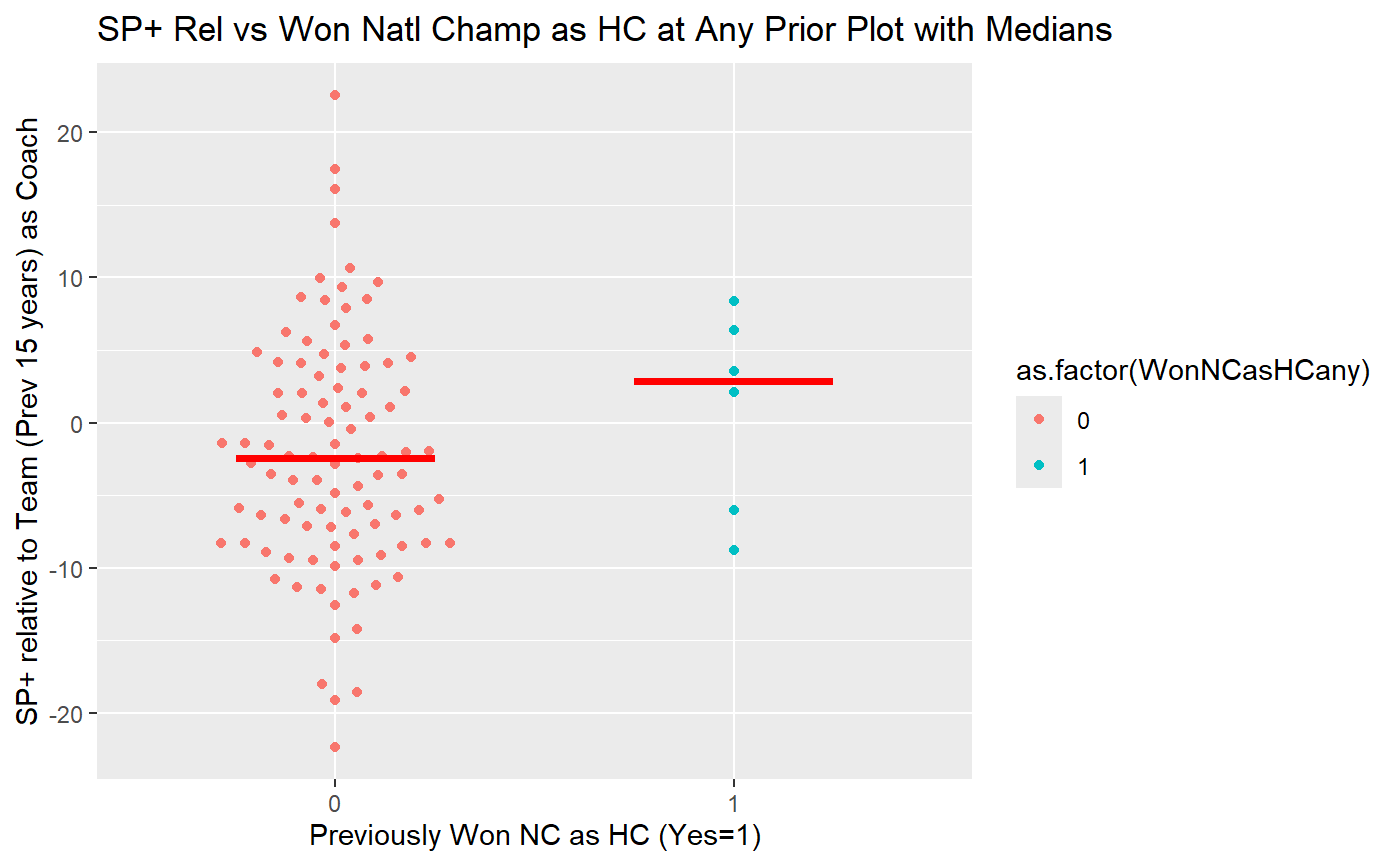}
    %\caption{
       \hspace*{.1in}
       {\footnotesize (b) Won Natl. Champ. as HC any level}
    \end{minipage}
\hfill 

\vspace{0.2in}

\begin{minipage}[htbp]{0.4\textwidth}
    \includegraphics[width=\textwidth]{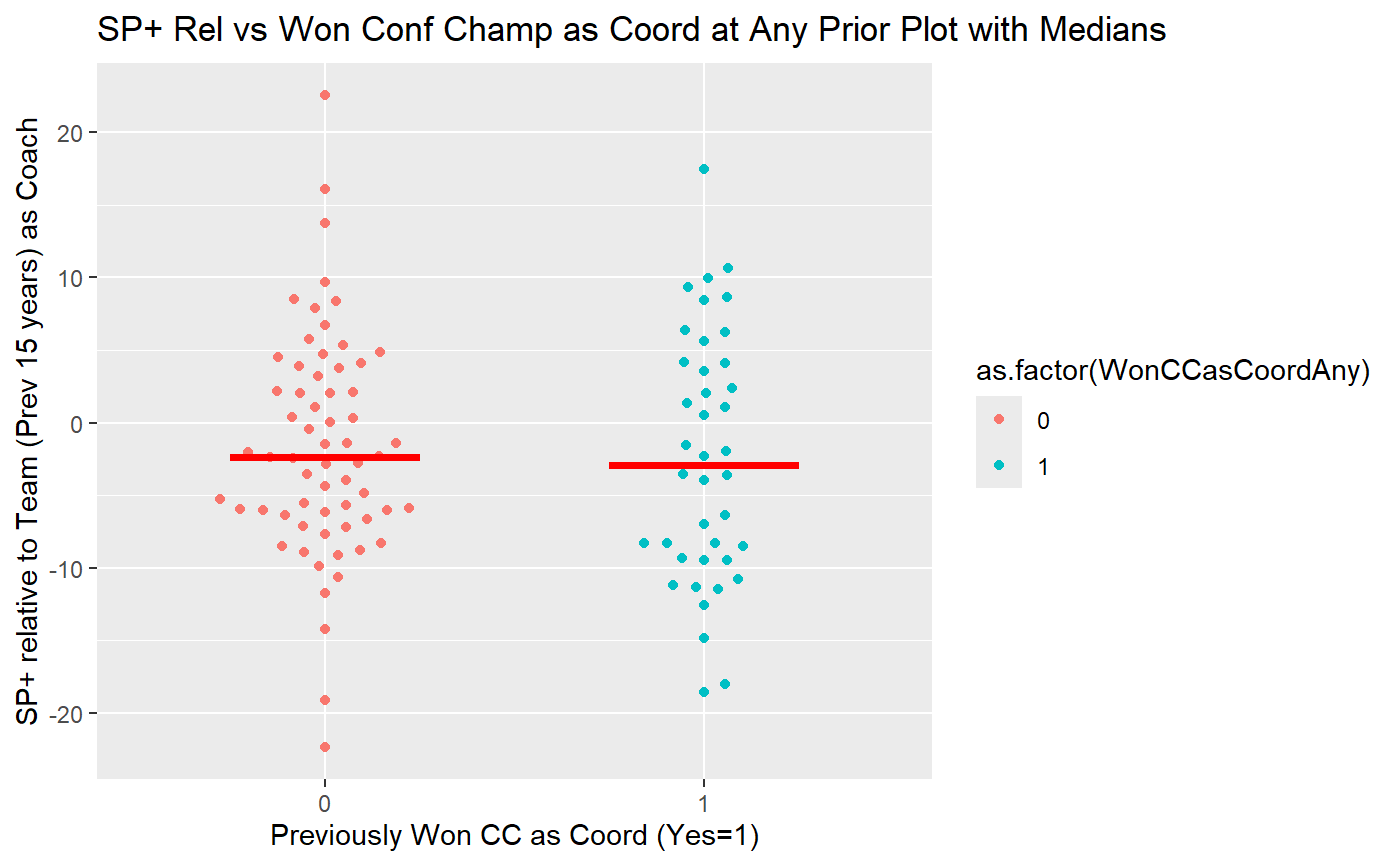}
    %\caption{
       \hspace*{.1in}
       {\footnotesize (c) Won Conf. Champ. as Coord any level}
    \end{minipage}
  %\hfill
 \begin{minipage}[htbp]{0.4\textwidth}
    \includegraphics[width=\textwidth]{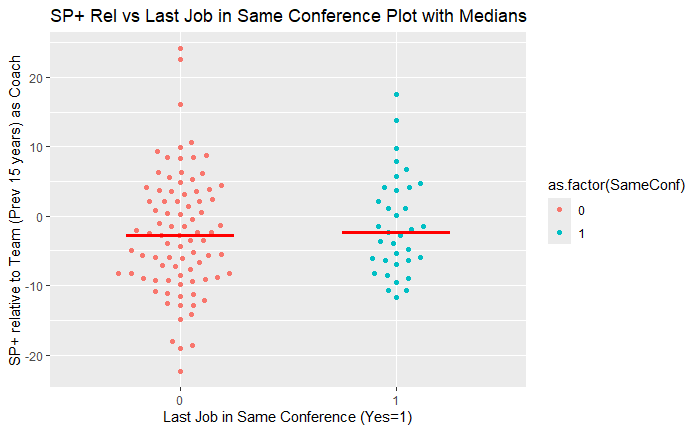}
    %\caption{
       \hspace*{.1in}
       {\footnotesize (d) Hired within same conference}
    \end{minipage}
\hfill

 \begin{minipage}[htbp]{0.4\textwidth}
    \includegraphics[width=\textwidth]{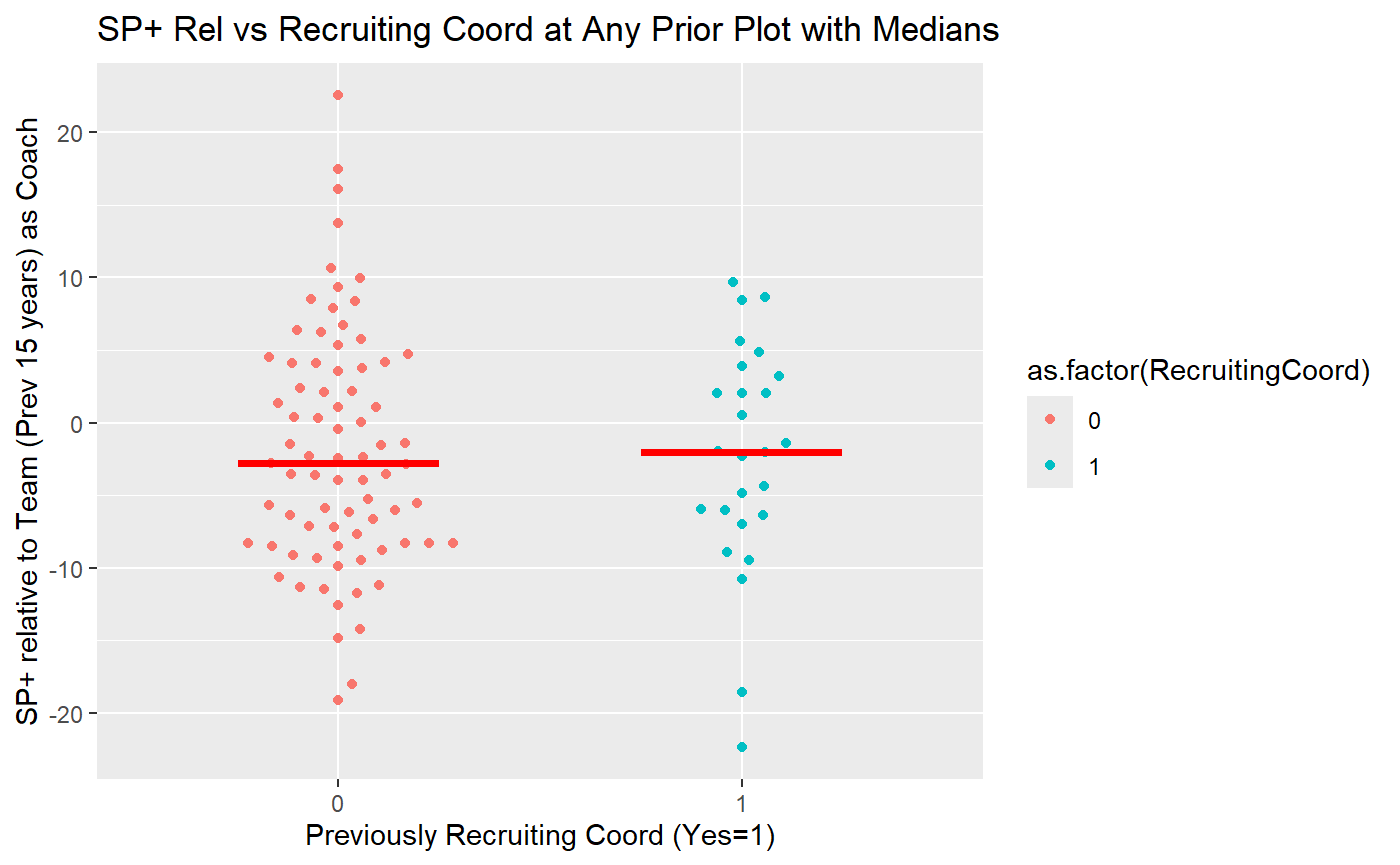}
    %\caption{
       \hspace*{.1in}
       {\footnotesize (e) Been Recruiting Coordinator previously}
    \end{minipage}
    
\end{figure*}

Next we will look at some of relationship between our variables and our measure of coaching 
performance, \textit{SP+ Rel}.  Table \ref{table:var} contains the list of predictor variables as well 
as their definitions.  Turning to the graphics, we see a sampling of these relationship in 
Figure \ref{fig:cat1}.  In that figure, we can see that medians (red lines) seem to 
indicate their are some differences between the categories for winning a national 
championship as a coordinator , Figure \ref{fig:cat1} (d), and for previous job
was as an OC, Figure \ref{fig:cat1} (g).  Otherwise the other variables do not 
seem to suggest there is much difference in the typical coaching performance 
for the different characteristics recorded in those subfigures.  More graphics
like these can be found for the other categorical predictors in Figures \ref{fig:append:cat}
and \ref{fig:append:cat2}.  In those graphics, Figure \ref{fig:append:cat} (d) 
suggests that having played in the NFL yields a higher median \textit{SP+ Rel} as
does having previously won a national championship as a coordinator, Figure \ref{fig:append:cat2} (b).  However, quite a bit fewer coaches have the value that is associated 
with the higher median.

Turning to numeric predictors, we can see the relationship between \textit{SP+ Rel} and
four of these predictors in Figure \ref{fig:numeric1}.  In Figure 2 (a), we
can see that there is not a relationship between number of years as a HC or 
Coordinator before being hired.  The three remaining subfigures, Figure 2 (b), (c),
(d), also show negative linear relationships between \textit{SP+ Rel} and
coach's age, teams performance in the recent past, 5 years, and team performance in the
previous 15 years.  More figures like this can be found in Figure \ref{fig:append:numeric} 
of the Appendix at the end of this report.  Noteworthy among those figure in the
Appendix is that there seems to be a positive linear relationship between
a hired coach's \textit{SP+ Rel} and their previous \textit{SP+ Rel} had they 
been a college HC previously.

\section{Model and Results}
The goal of this analysis is to build a predictive model for the success of a coaching hired and to identify the factors that impact that success.  We selected 29 predictors found in  Table \ref{table:var} that could be useful for predicting coaching success.  For our model formulation, we chose to build a regularized linear regression model for SP+ relative to past team performance, hereafter \textit{SP+ Rel}.  We used \texttt{cv.glmnet} as part of the glmnet library in R
\cite{glmnet}.  For our regularization, we considered both ridge and lasso regression, 
and we used 8-fold cross validated performance to determine the regularization/shrinkage parameter, $\lambda$, for our model.  We ran 20 different randomizations of 8-fold cross validation.  Consistently, across all these different randomizations of the folds, we found that the best lasso model had a smaller out of sample prediction error (via MSE) than the best ridge model.  To get the parameter coefficient estimates discussed below we fit the full model via a lasso with $\lambda$ set to the average of the 
shrinkage parameters that minimized the cross-validation error.  In fact the there was no variability in these shrinkage parameters across the 20 randomizations.  Because some 
of our predictors, eg sp\underline{~}prev\underline{~}rel, were available only if a coach
had previously been a college HC, we included whether or not a coach was previously an HC in
all of the models we considered.

Our final prediction equation is 

\begin{align}
\label{predictioneq}
\widehat{SP+ Rel} = -0.054
                    & + 0.493 \mbox{ Was College HC before at any level} \nonumber\\
                    & + 3.750 \mbox{ Previous job was as an OC}\nonumber \\
                    & -0.002 \mbox{ (Age - 50)}\nonumber \\
                    & -0.307 \mbox{ Hiring School's Team SP+ in 15 years before hiring}.
\end{align}
The largest factors in our model were whether or not the coach's previous job was as an
OC (PreviousOC), whether they had ever been a college coach previously 
(prevCollegeHC) and the teams performance over the last 15 years (team\underline{~}sp\underline{~}prev).  Having been hired from a OC position, is associated with about a $3.75$
point increase in a coach's \textit{SP+ Rel}.  While if a hire was previously 
a college football HC, the predicted increase in their team's performance increases
by nearly half a point.  Age is a factor in our model that is negatively correlated with coaching success though the effect is very small, about two-hundredth of a point per game for an additional 10 years of age.  This seems quite negligible.    For a newly hired 50 year old coach on a teams who's SP+ averaged
0 in the last 15 years with none of the other characteristics in the model, 
we expect their \textit{SP+ Rel} to be -0.054 or just slightly below what the
team had done in the previous 15 years.   
The last term in our prediction equation 
involves a team's average \textit{SP+ Rel} over the previous 15 years and that is
negatively associated with how a coaching hire's teams will perform.  We can 
interpret the value here of $-0.307$ as the expected decrease in a team's average
performance for every additional point of \textit{SP+ Rel} that the hiring team 
had over the previous 15 years.  This is a regressive effect, since SP+ for
an average team should be around zero, then for a new coaching hire
 good teams we would expect them to get worse
and  very good teams we would expect them to get much worse, while for
below average teams we would expect them to get better just by making the hire.  
 One possible explanation for this is a regression to the mean as team performance tends to move to the average of all teams, which for SP+ is around zero.  Another explanation is college football 
 has moved to an era of more parity and relative coaching performance is not immune 
 to moving with that trend.  

 The hires with the ten highest and ten lowest predicted \textit{SP+ Rel} are listed
 in Table \ref{table:topcoaches:pred}.  We can see that the model suggests that
 the hire with the highest predicted success was Rhett Lashlee at SMU.  Eight of
 the ten top predicted \textit{SP+ Rel} values had positive observed values.  
 Tony Elliott at Virginia and Jedd Fisch at Arizona were the exceptions.  Elliott
 has improved his Virginia teams and, if he stays, may yet get to positive territory.
 Fisch left Arizona after a successful second season.  Among the coaches
 predicted to have negative \textit{SP+ Rel}, eight of the ten are in or have
 finished in negative territory.  The exceptions are Dan Mullen whose team performance
 dropped significantly over his four years at Florida and Kirby Smart who has 
 impressively elevated Georgia to new heights.  Six of these ten coaches
 have already been fired with only Mullen, Smart, Fickell and Venables still employed
 at these schools.  Venables' Oklahoma 2025 team is having their best season
 with him as a HC and so there is hope for a turnaround there.  
 Looking at the bottom ten schools, most 
 of the schools there are considered to be `elite' or nearly so.  The model
 predicts that succeeding at these schools is hard partly because they are already
 at the upper tail of the distribution.  Overall the model accurately predicts 
 whether a coach will have a positive or negative \textit{SP+ Rel} 67\% of the time.
 Table \ref{table:truth}

\begin{table*}[h]
\centering
\caption{Top 10 \& Bottom 10 Coaching Hires by Predicted \textit{Sp+ Rel}}
\label{table:topcoaches:pred}
\begin{tabular}{|l||c|r|r|}
\hline
Name& Hiring School & \textit{SP+ Rel}&$\widehat{SP+ Rel}$\\
\hline
 Rhett Lashlee        &       SMU       & 17.8  &6.3\\
      Josh Heupel    &           UCF  &      22.6 & 5.7\\
 Major Applewhite    &       Houston  &       6.7 & 4.2\\
       Mike Gundy    &Oklahoma State  &      13.8 & 4.2\\
      Sonny Dykes    &           SMU  &      16.1 & 4.1\\
      Tony Elliott    &      Virginia  &      -5.6 & 3.5\\
     Mike Locksley    &      Maryland  &       0.4  &2.9\\
        Jedd Fisch    &       Arizona  &      -5.3  &2.7\\
      Joe Moorhead &Mississippi State  &       8.5  &2.7\\
    Lance Leipold  &          Kansas  &       8.5  &2.3\\
\hline
\hline
       Dan Mullen   &        Florida    &     1.1 &-5.7\\
      Mark Richt   &          Miami    &    -2.4 &-5.8\\
    Luke Fickell   &      Wisconsin    &   -15.0 &-5.9\\
      Tom Herman   &          Texas    &    -9.3 &-6.2\\
     Kirby Smart   &        Georgia    &     4.5 &-6.4\\
   Willie Taggart   &  Florida State    &   -19.1 &-6.4\\
     Jim McElwain    &       Florida     &   -8.3 &-6.6\\
      Brian Kelly    &           LSU     &   -6.0 &-6.7\\
   Brent Venables    &      Oklahoma     &   -8.9 &-6.8\\
       Ed Orgeron    &           LSU     &   -4.8 &-7.4\\
\hline
\end{tabular}
\end{table*}

The model we have explains about $24\%$ of the variability in the team's performance with a 
newly hired coach and the average difference between our model's prediction and the actual \textit{SP+ Rel}
of the coaches in these data is just over $7$ points.  In 2025 terms, that difference
is akin to the average difference in weekly performance between Texas A\&M and Missouri, or between USC and Nebraska or between Georgia Tech and NC State.  That is a large amount of variability and it
speaks to the noisiness of these data.  However, the model does well given that noisiness at predicting
who will have a \textit{SP+ Rel} above or below zero achieving 66\% accuracy on this metric.

While it is important to focus on the factors that do influence hiring success, it is worth mentioning those that do not given the factors that \textbf{are} in the model.  Having a previous connection to the school, being a previous head coach at any level, having NFL coaching experience, having experience as a recruiting coordinator, having played football in the NFL or at a P4 school seem to have no impact on our ability to predict how well a coach will do.  Winning a national championship as a HC at any level as well as being a Coordinator for a team that won a national championship or a conference championship does not predict coaching success.  Similarly having regional ties to the hiring school does not seem to be impactful and the same is true for moving within the same conference or leaving a P4 school.  A coach’s relative SP+ as a previous HC is also not predictive, as is their previous win percentage.  One possible explanation is that we are taking averages of previous coaching performance that misses changes in SP+ over a coaches times with a previous school, eg Matt Rhule at Temple and Lance Leipold at Buffalo.  Finally, the number of games coached as an HC before being hired is unimportant as a predictor as is the recent winning percentage of the team.

\section{Discussion}

Choosing a successful P4 college football coach is a difficult task.  In this paper we have 
collected data on recent P4 coaching hires and built a predictive model for their performance.  
The data we used was on recent P4 hires and we measured their performance using average 
SP+ relative to previous performance of the school’s teams, \textit{SP+ Rel}.   
As mentioned above \textit{SP+ Rel}
is  correlated with a coach's winning percentage and highly correlated with their SRS, the other
statistical measure of team performance that we considered.

Ultimately, we found that there are several factors that are predictive of the
performance of a college football head coaching hire.  
These factors include whether a hire had been a previous college coach,
whether the hire had previously won a conference championship as a coach, whether
their prior job was as an offensive coordinator, their age and the quality of the team
they were hired to lead over the previous 15 years.  Our prediction model does not
perform particularly well, only explaining just under a quarter of the variability in 
coaching performance. 

The 2020 season is among the years in these data.  The effects of COVID
were felt across college football in the form of shortened seasons and
empty stadiums.  Every school and every conference handled the situation
differently.  One of the choices we made as part of this analyis was that we 
would keep the 2020 season in our data.  Given the short tenure of some college
football head coaching hires, we felt having as much data as possible was
important.  

This study  measured the success of a coaching hire by how they did on average relative to
their team's previous 15 years.  That is not a perfect measure.  Team performance
fluctuates and trends, sometimes upward and sometimes downward.  Ideally when a new
coach is hired their teams trend upward as they recruit players that fit their system
and they become comfortable in their new job.  An average over a short period may miss
this.  That said, it is very difficult to generate a measure of coaching success that
captures all of the possible nuances.  Therefore, we chose to stick with an average
relative to past team performance.

\begin{figure*}[!t]
\centering
\caption{Additional plots comparing numeric predictors and \textit{SP+ Rel}}
\label{fig:append:numeric}
  %\hfill
 \begin{minipage}[htbp]{0.4\textwidth}
    \includegraphics[width=\textwidth]{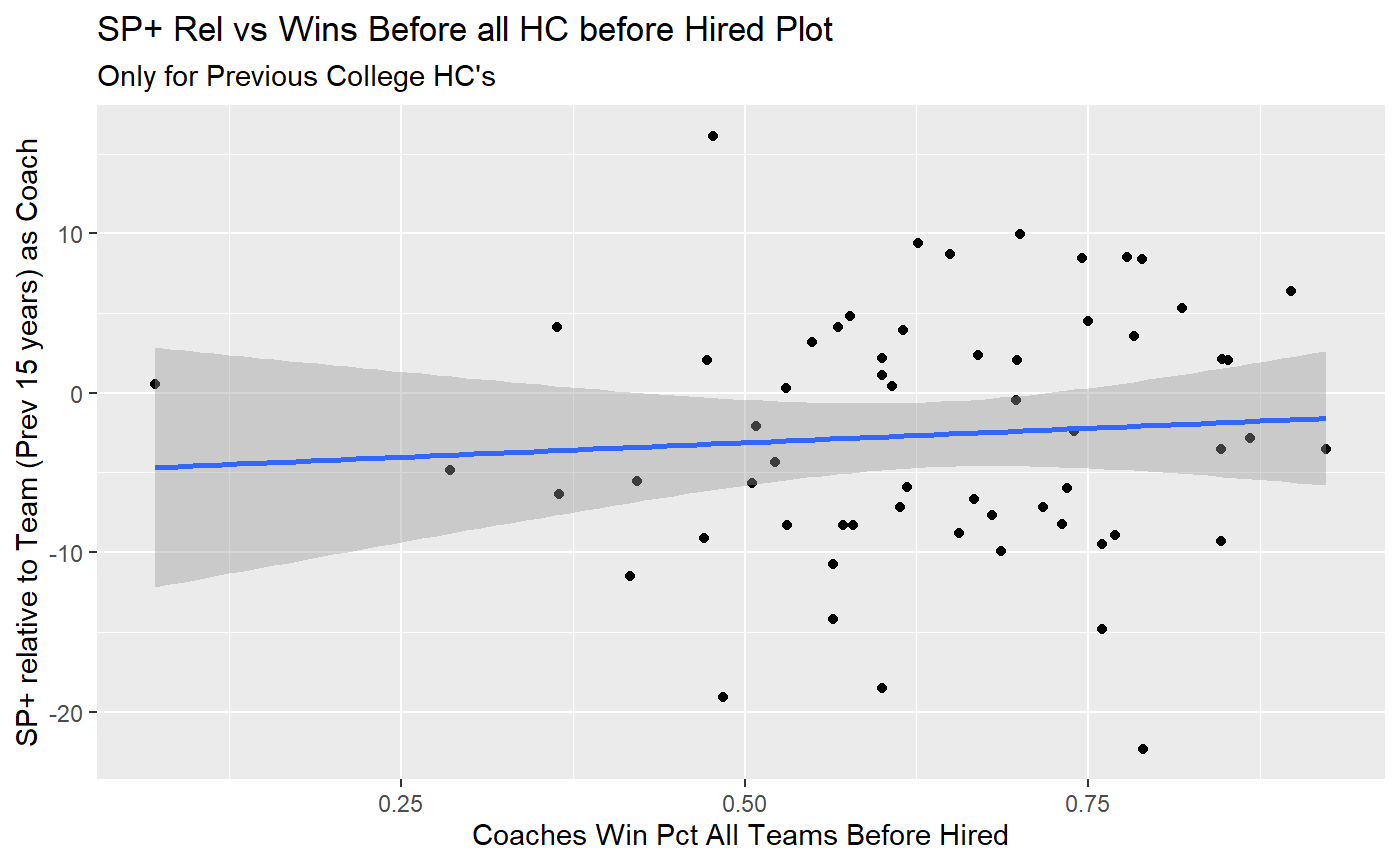}
    %\caption{
       \hspace*{.1in}
       {\footnotesize (a) All Past Win Pct as HC}
   
    \end{minipage}
\begin{minipage}[htbp]{0.4\textwidth}
    \includegraphics[width=\textwidth]{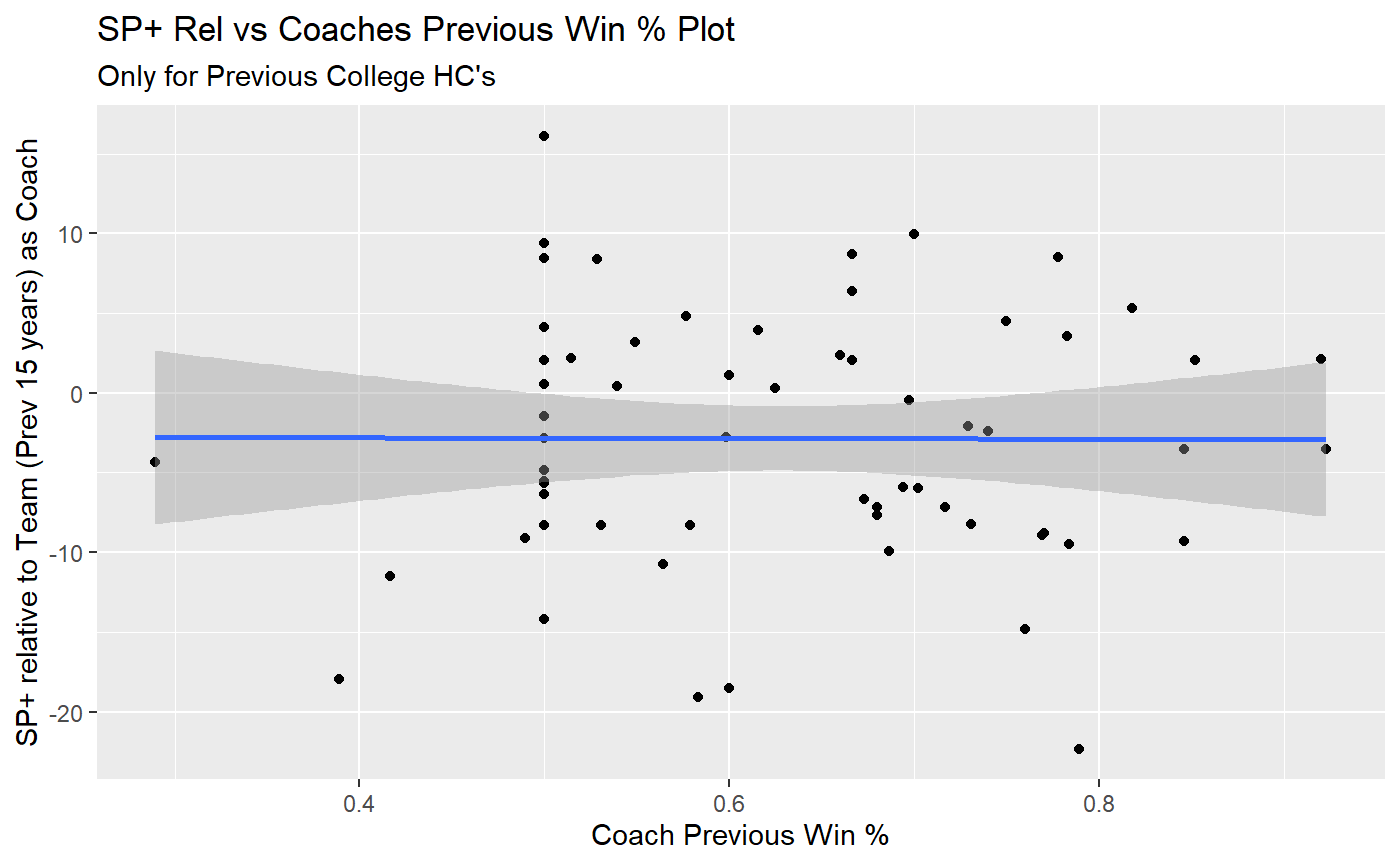}
    %\caption{
       \hspace*{.1in}
       {\footnotesize (b) Win Pct at Previous job if HC}
   
    \end{minipage}
\hfill

\vspace{0.2in}

\begin{minipage}[htbp]{0.4\textwidth}
    \includegraphics[width=\textwidth]{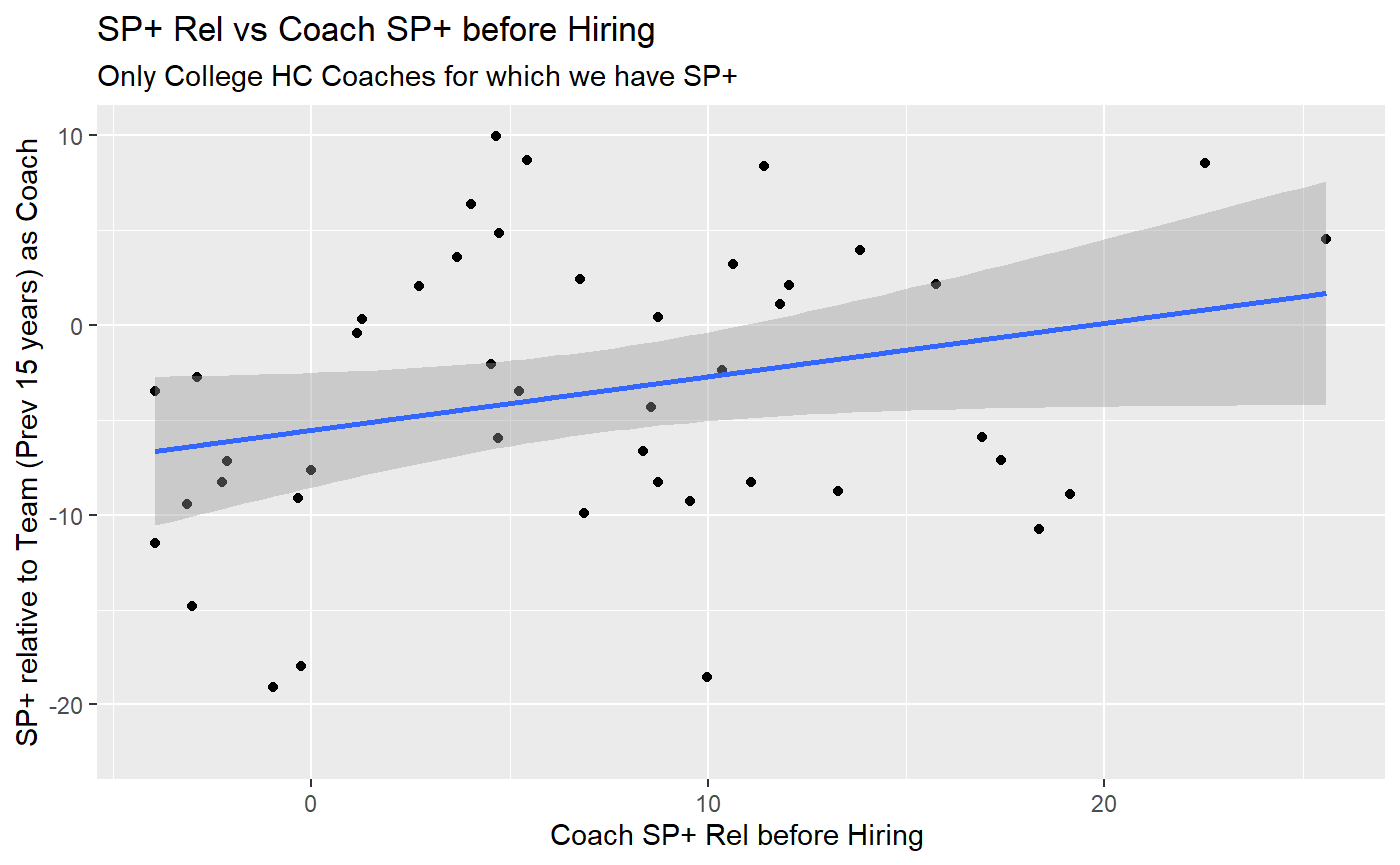}
    %\caption{
       \hspace*{.1in}
       {\footnotesize (c) Coach \textit{SP+ Rel} before hire}
    \end{minipage}
 \begin{minipage}[htbp]{0.4\textwidth}
    \includegraphics[width=\textwidth]{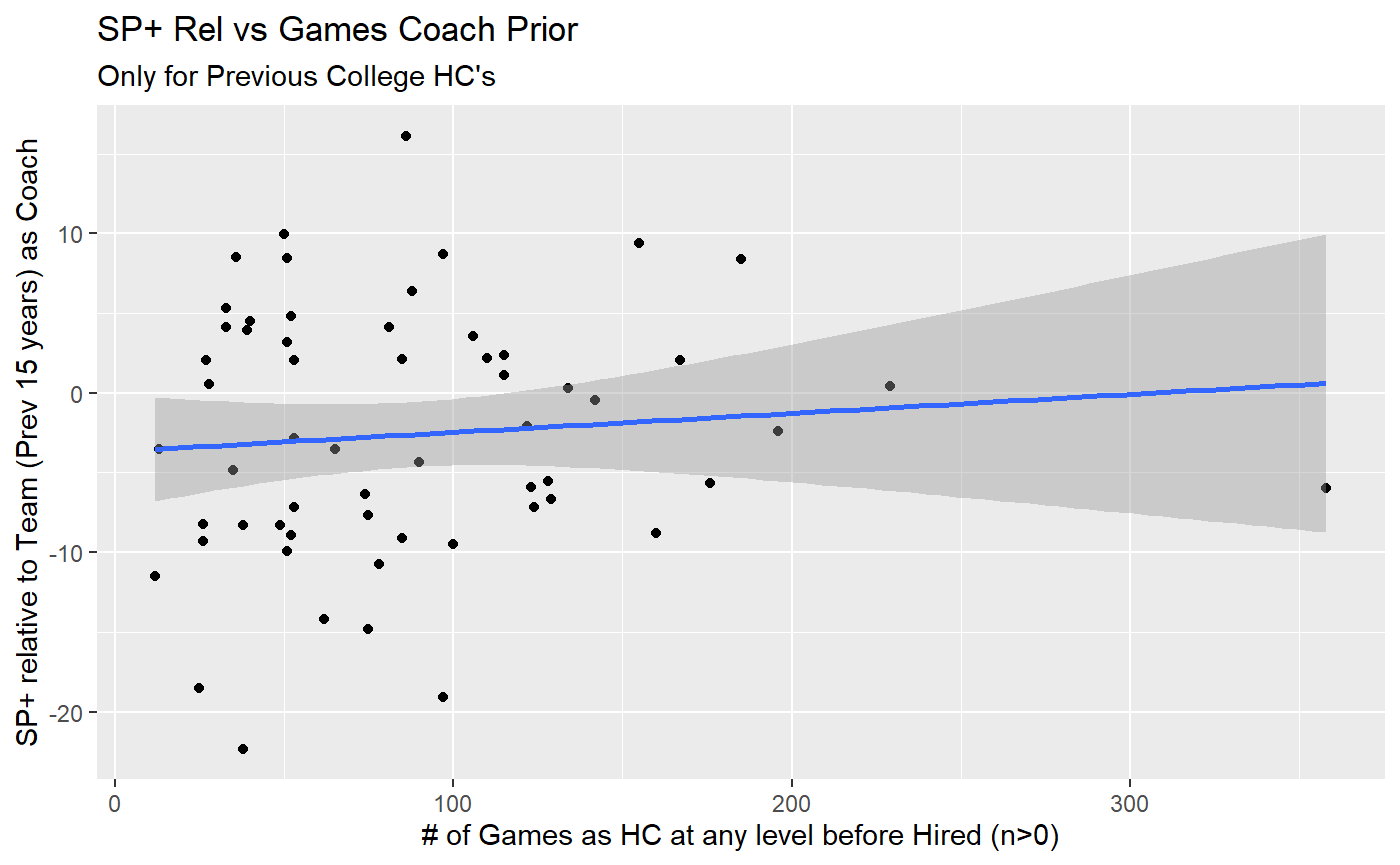}
    %\caption{
       \hspace*{.1in}
       {\footnotesize (d) Numb. of games as HC}
    
    \end{minipage}
  \hfill
 
\begin{minipage}[htbp]{0.4\textwidth}
    \includegraphics[width=\textwidth]{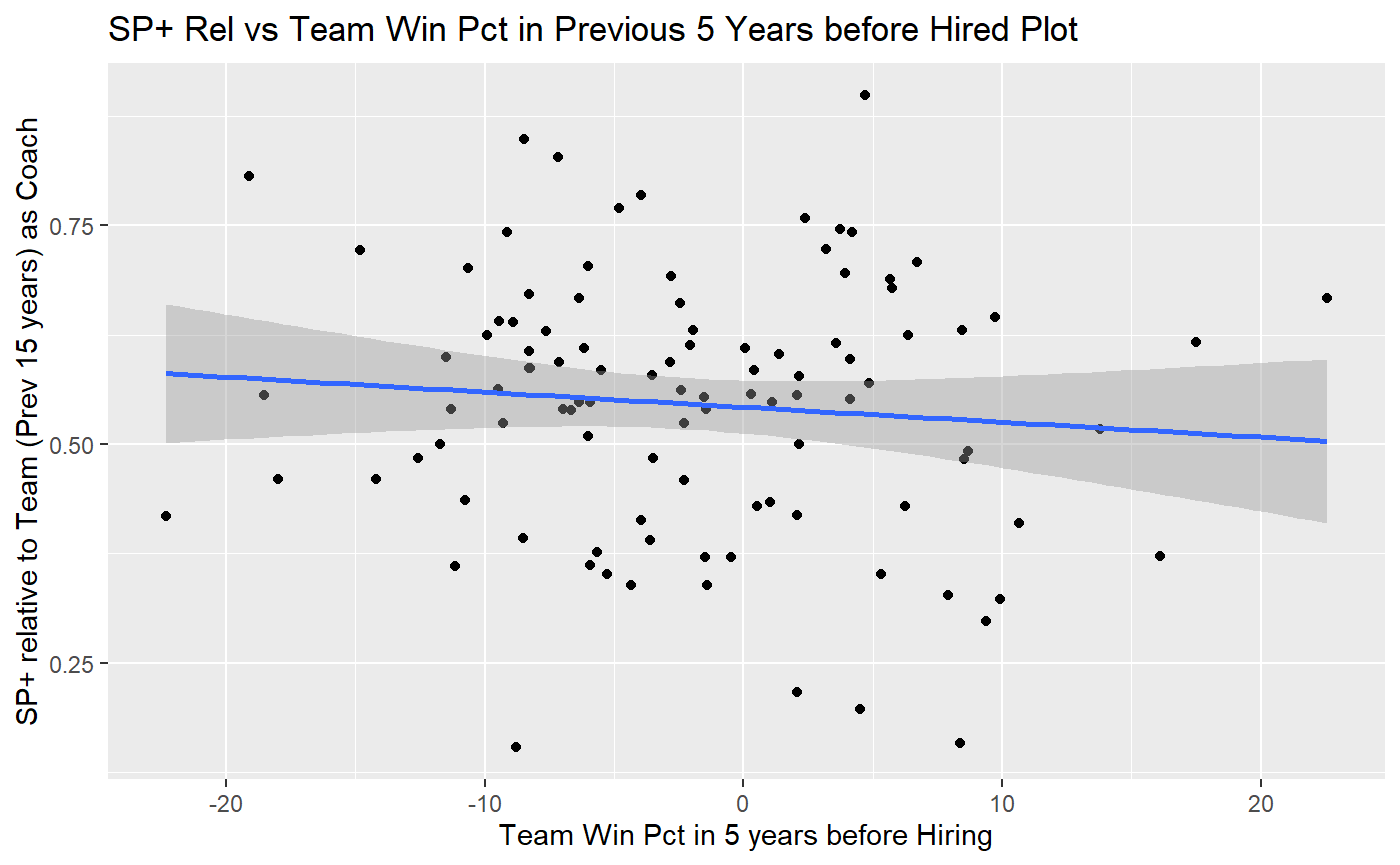}
    %\caption{
       \hspace*{.1in}
       {\footnotesize (d) Team Win Pct in last 5 years}
  
    \end{minipage}

\end{figure*}

Much has been made about the recent success of Curt Cignetti and he has done an
amazing job as the HC at Indiana.  So it is worth mentioning that Cignetti is 
not in the data we used for modelling because we have
less than two full years of him coaching at Indiana.  Regardless, Cignetti is an outlier.  It is
possible to regenerate all of the figures in this analysis with Cignetti and other coaches
hired after the 2023 season and it is very clear which observation corresponds to the
exceptional job Cignetti has done at Indiana.  That said, an outlier is not a trend and the
signal from these data is weak.   Hiring a coach is hard.  There are no magic attributes, at
least \textit{not} among the factors we investigated, that guarantee future success.  
The data as can be seen in Figures \ref{fig:cat1} through \ref{fig:append:numeric}
is very noisy and the trends, while tempting, are mild.  Great coaching are hard to 
find and many schools find themselves waiting for their
next great coach.

The code and the data for the analyses presented here can be found at \\
\href{https://github.com/schuckers/cfb_coach_analytics}{https://github.com/schuckers/cfb\underline{~}coach\underline{~}analytics}.

Thanks to the team of folks who put together \texttt{cfbfastr} for making their
data available in a usable and regularly updated manner and to Brian Macdonald (Yale) for
thoughtful discussions about this analysis.

\bibliographystyle{spmpsci}   
\bibliography{sports} 

\newpage
\section{Appendix}

\begin{table*}[h]
\centering
\caption{Worst Performing Coaching Hires by \textit{Sp+ Rel}.}
\label{table:bottomcoaches}
\begin{tabular}{|l||c|c|}
\hline
Name& Hiring School&\textit{SP+ Rel}\\
\hline
   Troy Taylor	&Stanford&	-22.3		\\
Willie Taggart	&Florida State&	-19.1		\\
Geoff Collins	&Georgia Tech&	-18.5		\\
Chad Morris	&Arkansas&	-18.0		\\
Luke Fickell	&Wisconsin&	-15.0		\\
Karl Dorrell&	Colorado&	-14.2		\\
Jeff Hafley	&Boston College&	-12.6		\\
Brent Pry	&Virginia Tech&	-11.8	\\
Ryan Walters	&Purdue&	-11.7		\\
Mel Tucker	&Michigan State&	-11.5		\\
\hline
\end{tabular}
\end{table*}

\begin{table*}[b]
\centering
\caption{List of variables used as predictors in this paper.  All values
are relative to the coach at the time they were hired.}
\label{table:var}
\begin{tabular}{|l||l|}
\hline
Variable Name & Variable Meaning\\
\hline\hline
prevCollegeHC& Were they previously a college head coach (HC)\\
\hline
PreviousHC& Was job when hired as a HC \\
\hline
PreviousOC& Was job when hired as an Offensive Coordinator(OC)\\
\hline
PreviousDC& Was job when hired as a Defensive Coordinator(DC)\\
\hline
Conf.LeavingP4 & Was job when hired at P4 school\\
\hline
SameConf& Was job when hired in same conference\\
\hline
AgeAtHiringMinus50& Age on Jan 1 of first season as HC\\
\hline
ConnectionToHiring.School& Had coach previously played or been employed at hiring school\\
\hline
NFLCoachingExperience& Had the coach every coached in the NFL\\
\hline
RecruitingCoord& Previously been a recruiting coordinator or not\\
\hline
PlayedNFL& Had coach played in the NFL\\
\hline
Played.Power.4& Had coach played at a Power 4 school\\
\hline
WonNCasHCany& Had coach won a National Championship (NC) as HC at any level(college)\\
\hline
WonCCasHCany& Had coach won a Conference Championship (CC) as HC at any level (college)\\
\hline
WonNCasCoordAny& Had coach won a NC as a OC or DC or AHC at any level (college)\\
\hline
WonCCasCoordAny& Had coach won a CC as a OC or DC or AHC at any level (college)\\
\hline
SideO& Is coach's primary background on Offense\\
\hline
LeavingNFL& Was previous job in the NFL\\
\hline
Regional.Ties& Had coach previously worked within 500 miles of hiring school\\
\hline
numb\underline{~}years\underline{~}Coord\underline{~}adj & Number of years as OC/DC/AHC/HC above average of 11.8years\\
\hline
numb\underline{~}years\underline{~}Coord\underline{~}HCadj2 & 
Numbers of years as OC/DC/AHC/HC different from 11.8 squared\\
\hline
PromoteWithin& Coach was promoted to HC from other position at same school\\
\hline
AllPriorWinPct& Win Percent as Coach at all levels before being hired\\
\hline
winpct\underline{~}prev\underline{~}Rel& Win percent as HC at previous job\\
\hline
team\underline{~}sp\underline{~}prev& Hiring school's SP+ in  15 years before hire\\
\hline
team\underline{~}sp\underline{~}recent& Hiring school's SP+ in 5 years before hire\\
\hline
GamesAsHCPrior& Number of games as a HC before being hired\\
\hline
team\underline{~}winpct\underline{~}recent& Hiring School's win percentage in the last 5 years\\
\hline
sp\underline{~}prev\underline{~}rel& Previous \textit{SP+ Rel} in HC role\\
\hline

\end{tabular}
\end{table*}

\begin{table*}[h]
\centering
\caption{Truth Table for \textit{SP+ Rel}}
\label{table:truth}

\begin{tabular}{ll||r|r}
&&\multicolumn{2}{c}{actual}\\
&&\textit{SP+ Rel}$<0$& \textit{SP+ Rel}$>0$\\[5pt]
\hline\hline
predicted&$\widehat{SP+ Rel}<0$&51&23\\[5pt]
\hline
&$\widehat{SP+ Rel}>0$&11&18\\[5pt]
\end{tabular}
\end{table*}

\end{document}